\documentclass[aps,prc,twocolumn,showpacs,superscriptaddress,footinbib]{revtex4-1}

\usepackage{graphicx}
\usepackage{dcolumn}
\usepackage{bm}
\usepackage{enumerate}
\usepackage{amsmath}

\usepackage[normalem]{ulem}  

\begin{document}


\title{Spherical-like spectra for the description of the normal states of $^{108-120}$Cd in the SU3-IBM and the $Q_{2_{1}^{+}}$ anomaly}

\author{Tao Wang}
\email{suiyueqiaoqiao@163.com}
\affiliation{College of Physics, Tonghua Normal University, Tonghua 134000, People's Republic of China}

\author{Xin Chen}
\affiliation{Department of Physics, Liaoning Normal University, Dalian 116029, People's Republic of China}

\author{Yu Zhang}
\email{dlzhangyuphysics@163.com}
\affiliation{Department of Physics, Liaoning Normal University, Dalian 116029, People's Republic of China}

\date{\today}

\begin{abstract}
\textbf{Abstract:} The Cd puzzle implies that the phonon excitation of a spherical nucleus should be questioned and refuted. For understanding this spherical-like $\gamma$-soft mode newly found experimentally, a possible answer was proposed recently in the SU3-IBM. In this paper, the evolutions of the normal states in $^{108-120}$Cd are investigated and compared with the experimental results. For better explaining the nearly zero B(E2) values between the $0_{2}^{+}$ state and the $2_{1}^{+}$ state, except for the SU(3) second-order and third-order Casimir operators, other SU(3) higher-order interactions are also considered in detail. It can be found that the results of theoretical fitting and experimental data agree well with simple parameter selection. The spherical-like spectra really exist. The deficiency may come from the lack of configuration mixing. The realistic spectra characteristics of the spherical-like spectra are found for $^{118,120}$Cd and the electric quadrupole moments of the $2_{1}^{+}$ state are predicted. The $Q_{2_{1}^{+}}$ anomaly in $^{108-116}$Cd are also discussed.

\end{abstract}

\maketitle

\section{Introduction}

\label{intro}

\textbf{The basic assumption of nuclear structure is that nuclei with magic numbers are spherical \cite{Otsuka22,Wood22,Mayer55}. When the valence nucleons increase, spherical vibrational excitation seems to appear inevitably for an even-even nucleus where the residual two-body interaction should appear \cite{Wood22}. This excitation shows a nearly equidistant spectrum created by the quadrupole phonon. At the two-phonon excitation level, there are three degenerate states with angular momentum $L=4,2,0$ while at the three-phonon level, there are five degenerate states with $L=6,4,3,2,0$ \cite{Wood18}.}

\textbf{The interacting boson model (IBM) provide a simple but effective way to study the collectivity in the nuclei \cite{Iachello87,Casten10,Wang08}. Recently the extended interacting boson model with SU(3) higher-order interactions (SU3-IBM) was proposed \cite{wang22}, and one spherical-like $\gamma$-soft rotation was found in this model to explain the Cd puzzle \cite{Wood11,Wood16,Wood18}. The energy of the $0_{3}^{+}$ state is nearly twice the one of the $0_{2}^{+}$ state, so it does not appear near the $6^{+},4^{+},3^{+},2^{+}$ states. The $0_{3}^{+}$ state is expelled to a higher level. (see the bottom graph of Fig. 2 in Ref. \cite{wang})}

\textbf{This new model can be seen as an IBM realization of the SU(3) symmetry viewpoints adopted in the SU(3) shell models \cite{Elliott1,Elliott2,Harvey,Draayer1,Draayer2,Draayer3,Draayer4,Draayer5,Elliott83}, some of which were discussed with the pseudo-SU(3) symmetry \cite{Arima69,Hecht69}. Recently proxy-SU(3) symmetry was found by Bonatsos \emph{et al.} \cite{Bonatsos17,Bonatsos23}, which investigates the prolate-oblate shape phase transition within only the SU(3) symmetry. Higher-order interactions are needed in the IBM \cite{Isacker81} to include the triaxiality \cite{Davydov58,Wood04}.
SU(3) higher-order interactions are related to the SU(3) mapping of the rigid triaxial rotor \cite{Draayer87,Draayer881,Draater882}, which were also investigated to release the degeneracy of the $\beta$-band and the $\gamma$-band in the SU(3) limit \cite{Isacker85}, and to realize the rigid triaxial rotor \cite{Isacker00,zhang14}.}

\textbf{The spherical-like spectra in the Cd puzzle \cite{Wood11,Wood16,Wood18} has been only found in this SU3-IBM \cite{wang22}. B(E2) anomaly can be only explained by the SU3-IBM \cite{Wang20,Zhang22,Zhang24}, which can not be described by other nuclear theories \cite{166W,168Os,170Os,172Pt} and the O(6) symmetry \cite{wangtao}. Prolate-oblate shape phase transition was also found to evolve with an asymmetric way in this model, which is more consistent with the way the actual nuclei evolve in the Hf-Hg region \cite{wang23}. This model can also explain the energy spectra of $^{196}$Pt at a better level \cite{wang,zhou24} and also the $^{82}$Kr \cite{zhou23}. The SU3-IBM describes the oblate shape with the SU(3) third-order Casimir operator \cite{Fortunato11,Zhang12}, which predicts a boson number odd-even effect \cite{Zhang12}. This effect has been found recently in $^{196-204}$Hg, which conclusively proves the validity of this new model \cite{Wang24}. }

\textbf{Recently, Otsuka \emph{et al.} argue that nuclei previously thought to be prolate ellipsoid, such as $^{166}$Er, are actually of a triaxial shape \cite{Otsuka19,Otsuka21,Otsuka22,Otsuka}. This result is attractive and supports the SU3-IBM because it is very easy to handle triaxial deformations.}

\textbf{It is necessary to further investigate the Cd puzzle \cite{Wood11,Wood16,Wood18} with the SU3-IBM. In previous paper \cite{wang22}, only the SU(3) second-order and third-order Casimir operators were considered. In this paper, we further investigate other higher-order interactions to explain the nearly zero B(E2) values between the $0_{2}^{+}$ state and the $2_{1}^{+}$ state. The coupling strength between the normal states and the intruder states in $^{108-116}$Cd is weak \cite{Garrett12}. However, for $^{118,120}$Cd, we find that the coupling between the normal states and the intruder states can be ignored. Thus the spectra in the $^{118,120}$Cd can be explained without configuration mixing calculation (see TABLE I). Similar spherical-like spectra may be also found in the Pd nuclei \cite{Sona22,Yates}. Finally, for the first time, the anomalous evolutional behavior of the quadrupole moments of the $2_{1}^{+}$ state in $^{108-116}$Cd can be discussed by the theory, implying that the discussions are meaningful.}

\begin{table}[!ht]
    \centering
    \caption{Main viewpoints adopted in this paper.}
    \begin{tabular}{llllll}
    \hline
        Cd nuclei~~~~& coupling strength~~~~ & configuration mixing  \\ \hline
        $^{108-116}$Cd & weak   & needed           \\
        $^{118,120}$Cd  & ignored  & no   \\ \hline
       \end{tabular}
    \label{dsafdsfdfd}
\end{table}

\begin{figure}[tbh]
\includegraphics[scale=0.32]{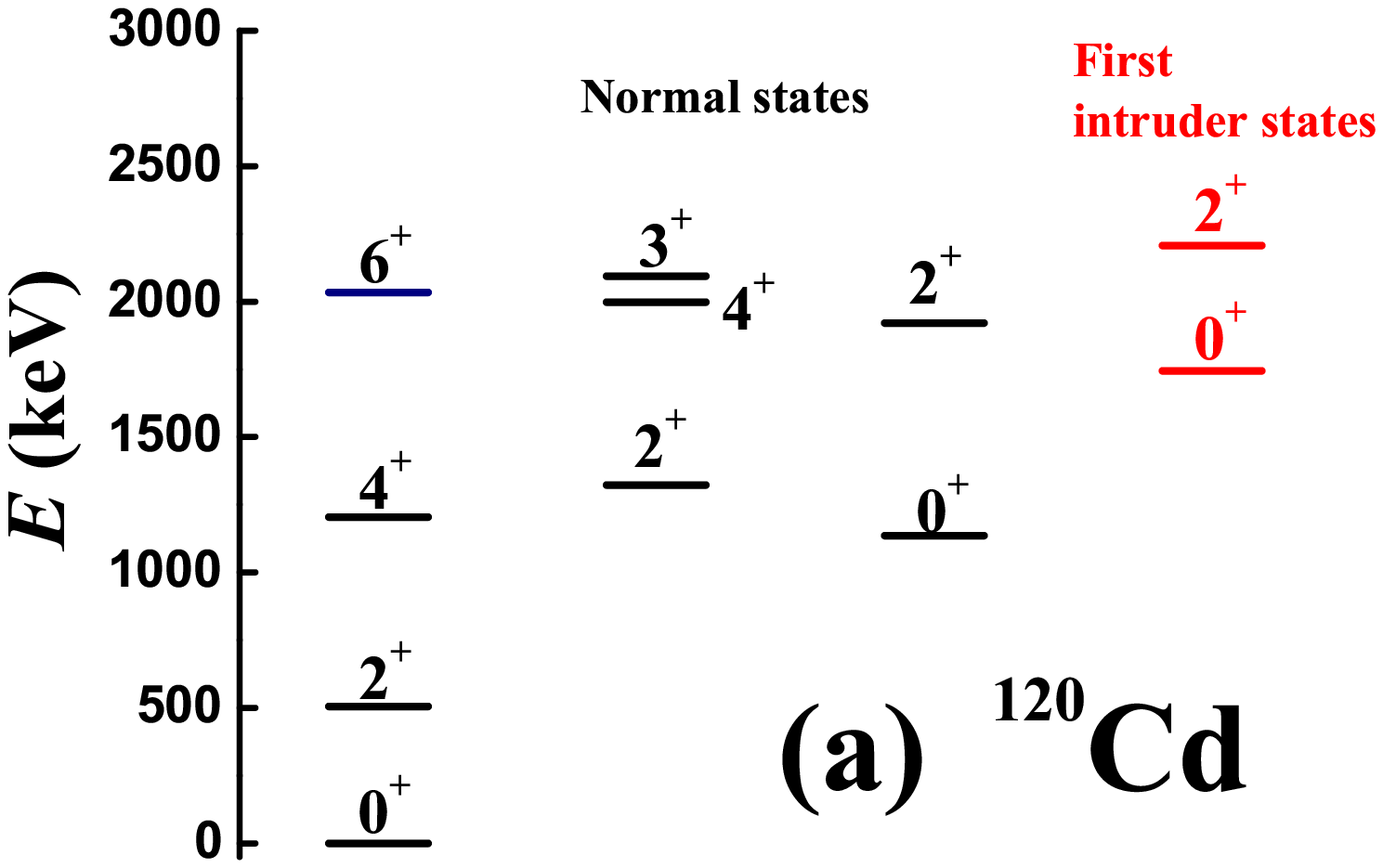}
\includegraphics[scale=0.32]{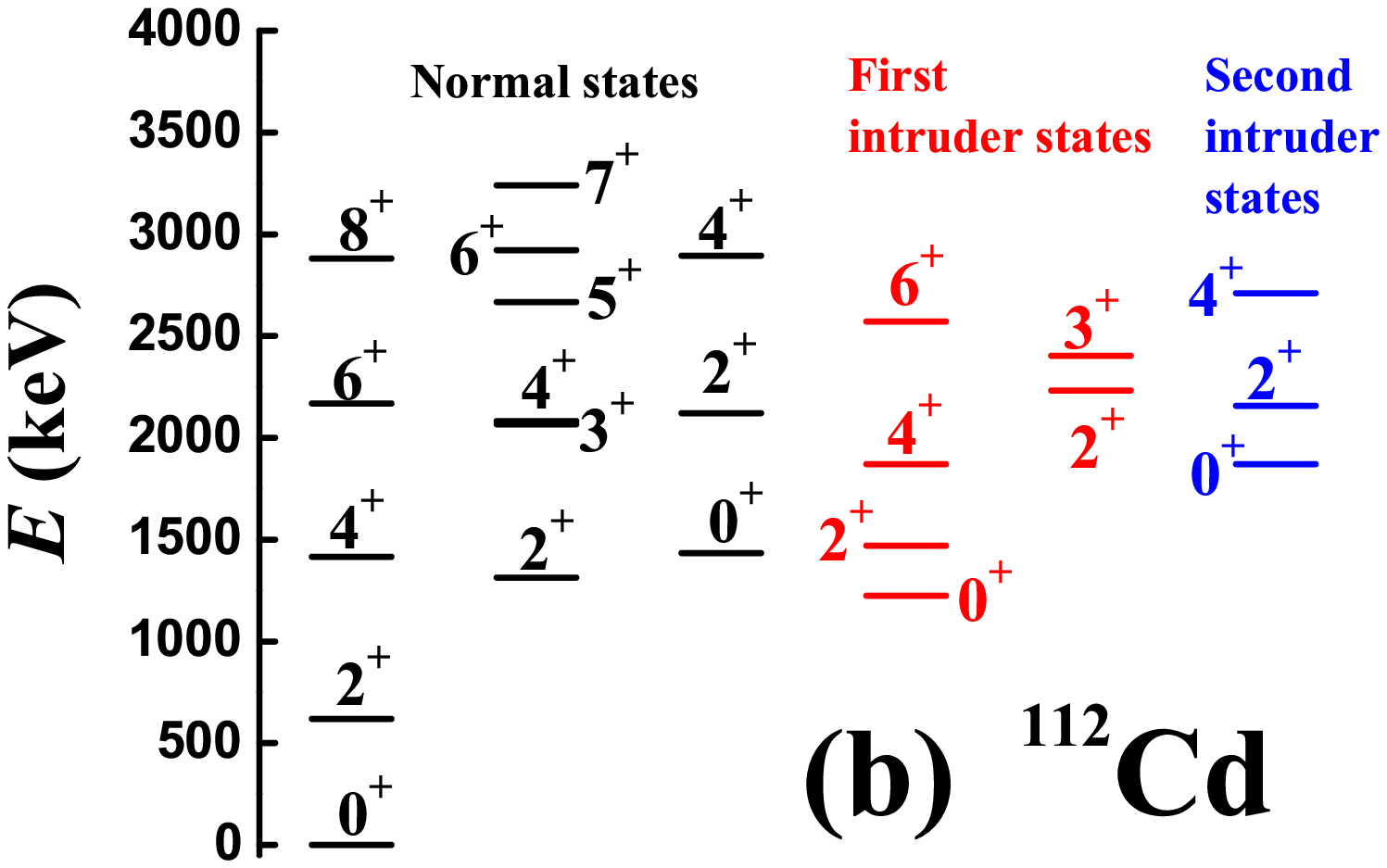}
\caption{Partial \textbf{experimental} low-lying spectra for (a) $^{120}$Cd and (b) $^{112}$Cd. The key observation is that there is in fact no the $0^{+}$ state at the three-phonon level.}
\end{figure}

\begin{figure*}[!htb]
	\resizebox{1.0\textwidth}{!}{%
		\includegraphics{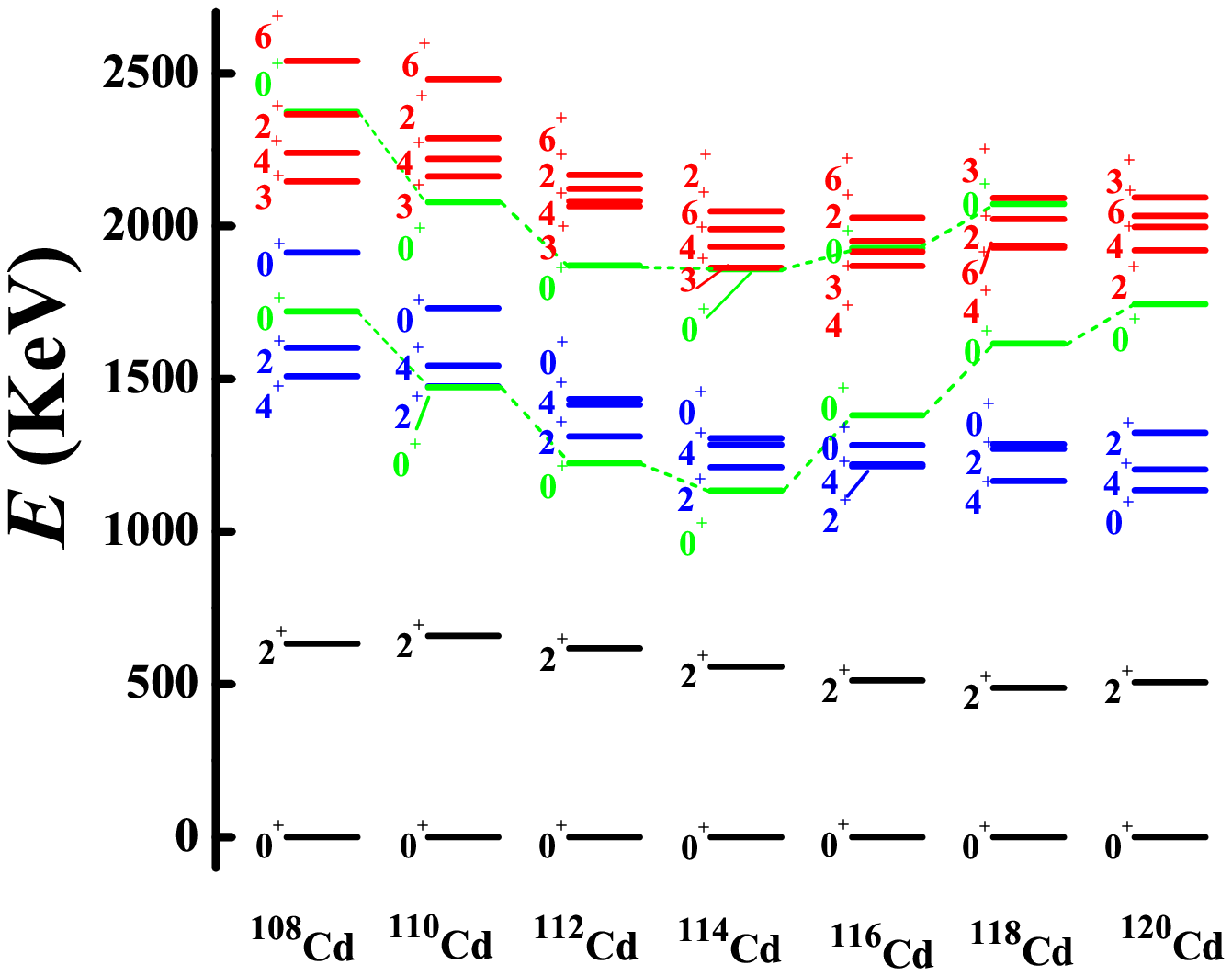}
	}
	\caption{\textbf{Experimental} systematics of the even-even Cd nuclei from 108 to 120. }
\end{figure*}

\section{Determining the normal states of $^{108-120}$Cd}

\textbf{The Cd puzzle is one of the difficulties discovered in nuclear structure physics \cite{Wood11,Wood16}. For similar phenomenon is also found in other nuclei, such as Te, Pd \emph{et al.} \cite{Wood18}, here it is called spherical nucleus puzzle \cite{wang22}. It was established long ago that the shape coexistence \cite{Wood11,Garrett22,Bonatsos,Otsuka24} exists in the Cd isotopes \cite{Cohen60,Meyer77}. It was also found that the electromagnetic transitions among the normal states are different from the phonon excitation modes. To explain this inconsistency, the researchers hypothesized that the coupling between the normal states and the intruder states is strong because they believed that the Cd nuclei are spherical \cite{Heyde82,Kern93,Jolie93}.}

\textbf{However new experiments do not support this strong coupling viewpoint \cite{Garrett08,Garrett10}. The coupling strength was found to be weak \cite{Garrett12}. This has caused a lot of confusions. If the coupling is weak, the normal states are not the evidence of a spherical nucleus (the Cd puzzle). From the electromagnetic transitions, it is more like a $\gamma$-soft rotor \cite{Garrett12}. Thus the experimental data revealed a new spherical-like $\gamma$-soft rotational mode.}

\textbf{In Ref. \cite{Batchelder12} it was experimentally found that $^{120}$Cd has no $0_{3}^{+}$ state near the three-phonon level, see Fig. 1 (a). The $0_{3}^{+}$ state is the bandhead of the first intruder states (red band). Thus for the normal states, the $0^{+}$ state at the three-phonon level disappears, and the four states with $L=6^{+},4^{+},3^{+},2^{+}$ are left. This feature can be also confirmed in $^{112}$Cd experimentally \cite{Garrett19,Garrett20}, see Fig. 1 (b). The $0^{+}$ state (blue band) at the three-phonon level was verified as the bandhead of the second intruder states. Thus at the three-phonon level in $^{112}$Cd, the $0^{+}$ state does not exist at all. The normal states in the Cd nuclei are in fact the black ones in Fig. 1.}

\textbf{More importantly, the two $0^{+}$ states (green color) in the evolutions of $^{^{108-120}}$Cd appear with obvious parabolic shapes, see the experimental systematics in Fig. 2. This is a typical feature of shape coexistence \cite{Wood11,Garrett22,Bonatsos,Otsuka24}. Obviously, the four $6^{+},4^{+},3^{+},2^{+}$  states evolve in a very different way from the $0_{3}^{+}$ state. In Fig. 2, it is obvious that the $0^{+}$ state (blue color) evolves in the same way as the ones of the two $4_{1}^{+}, 2_{2}^{+}$ states, which does not show the parabolic feature. Thus the normal states present a new spherical-like $\gamma$-soft spectrum. It is a $\gamma$-soft rotor, but not an O(6) nucleus. There are two key differences: (1) there is a $0_{2}^{+}$ state near the $4_{1}^{+}$ and $2_{2}^{+}$ states (spherical-like), and (2) there is a $2_{3}^{+}$ state near the $6_{1}^{+}$, $4_{2}^{+}$, $3_{1}^{+}$ states, and no the $0_{3}^{+}$ state.}

\textbf{The Cd puzzle was also studied with other perspectives \cite{Leviatan18,Leviatan23,Garrett19,Garrett20}, in which they don't regard the normal states in the Cd nuclei as a new collective mode. The properties of the Cd nuclei were also discussed in \cite{Nomura18,Nomura22}, which provides a
reasonable qualitative description of the experimental energy levels and transition rates.}

\begin{figure}[tbh]
\includegraphics[scale=0.30]{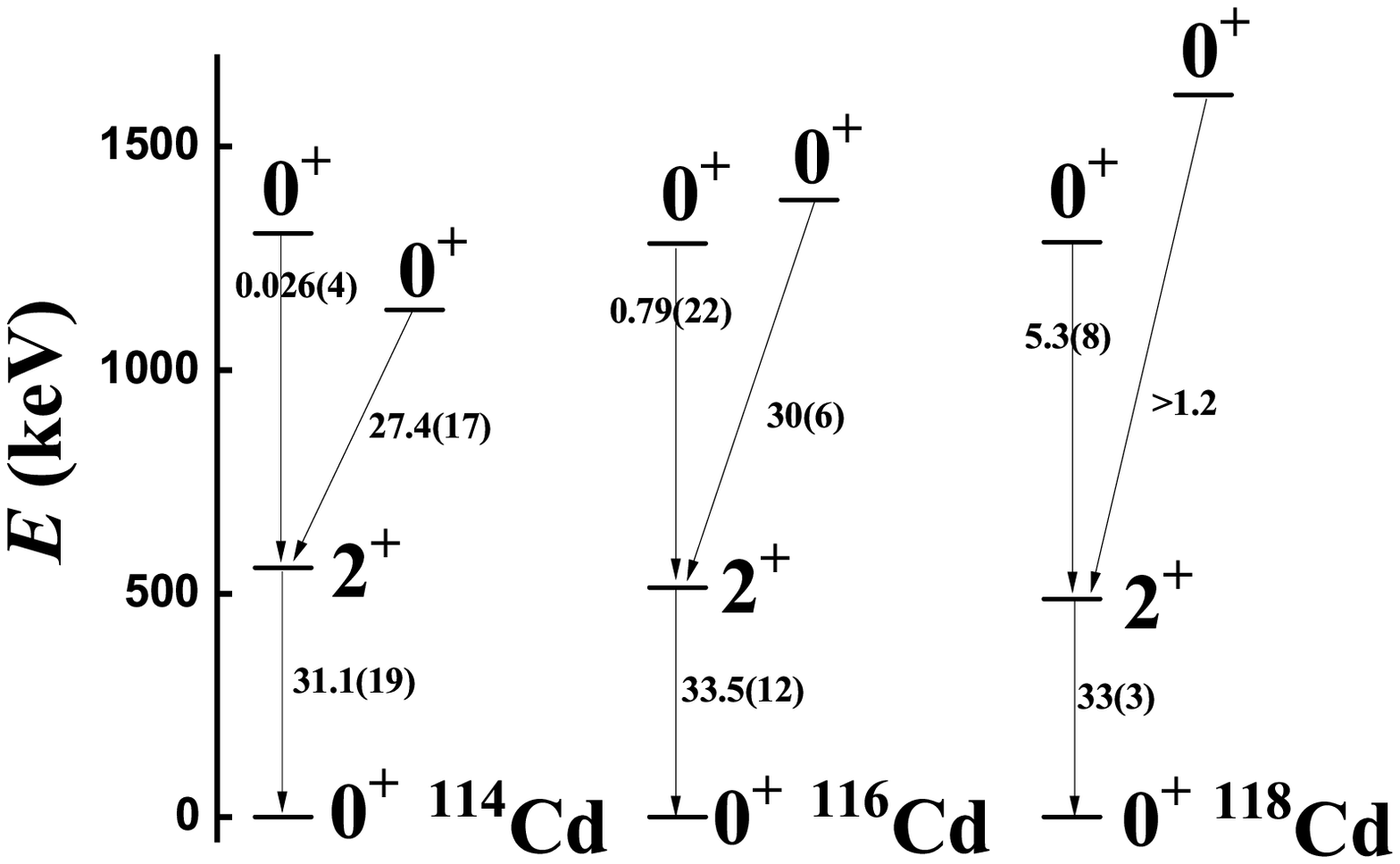}
\caption{\textbf{Experimental} B(E2) values (W. u.) among the first $2^{+}$ state and the lowest three $0^{+}$ states in $^{114-118}$Cd.}
\end{figure}

\textbf{A key point should be stressed here, see Fig. 3. From $^{114,116}$Cd to $^{118}$Cd, the experimenta level position of the intruder $0^{+}$ state moves up, and for $^{118}$Cd, the coupling between the normal states and the intruder states can be ignored. In Fig. 3, it can be found that the B(E2) value between the $0^{+}$ state and the $2^{+}$ state in the normal states changes rapidly from nearly zero to around 5.3 W. u., which implies that the nearly zero B(E2) value in $^{114,116}$Cd is partly induced by the configuration mixing. The normal states of $^{120}$Cd are the highest, so the coupling can be also ignored, see Fig. 2. Thus we suggest that the normal states of $^{118,120}$Cd are typical of the new spherical-like $\gamma$-soft rotation.}

\begin{figure}[tbh]
\includegraphics[scale=0.32]{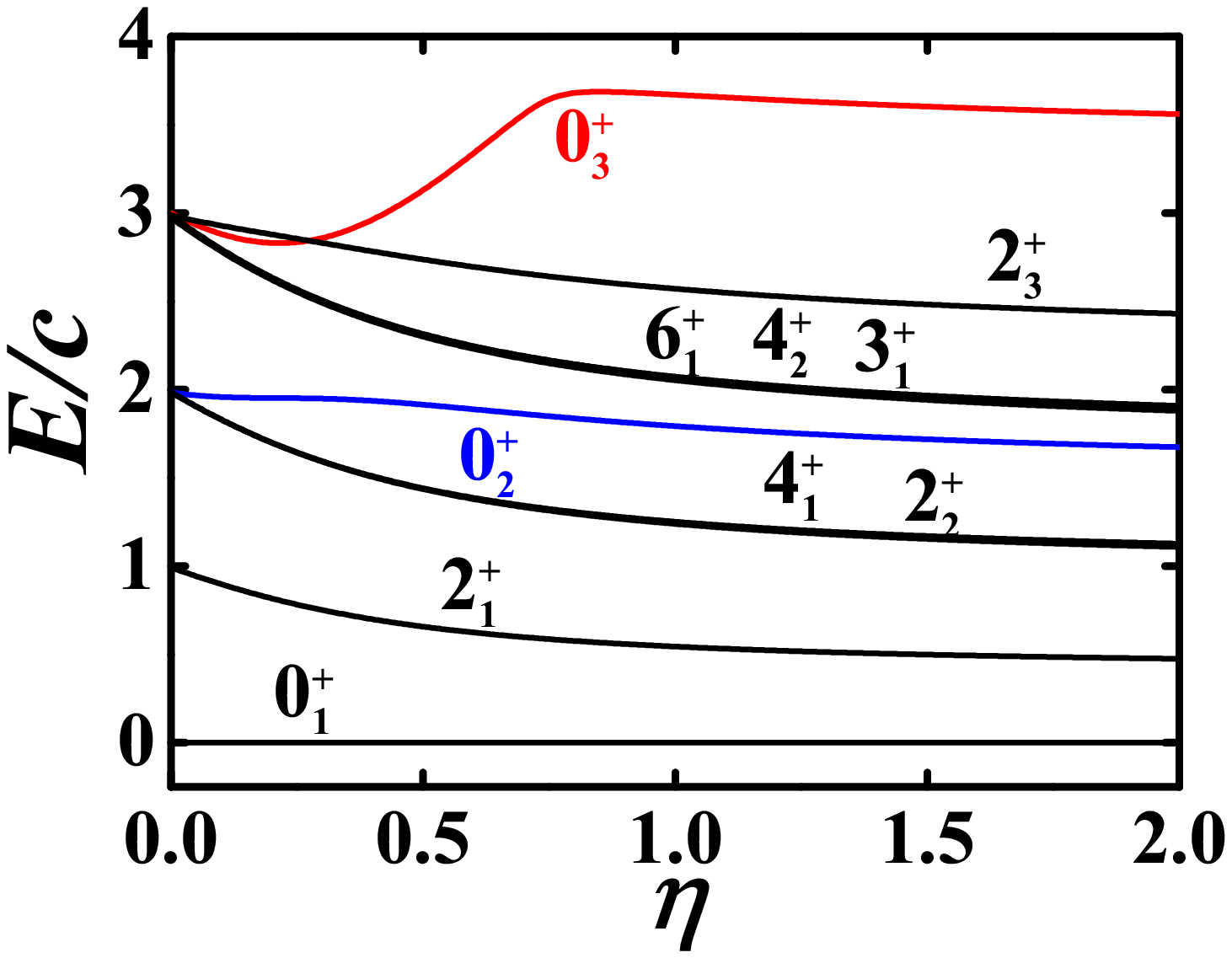}
\includegraphics[scale=0.32]{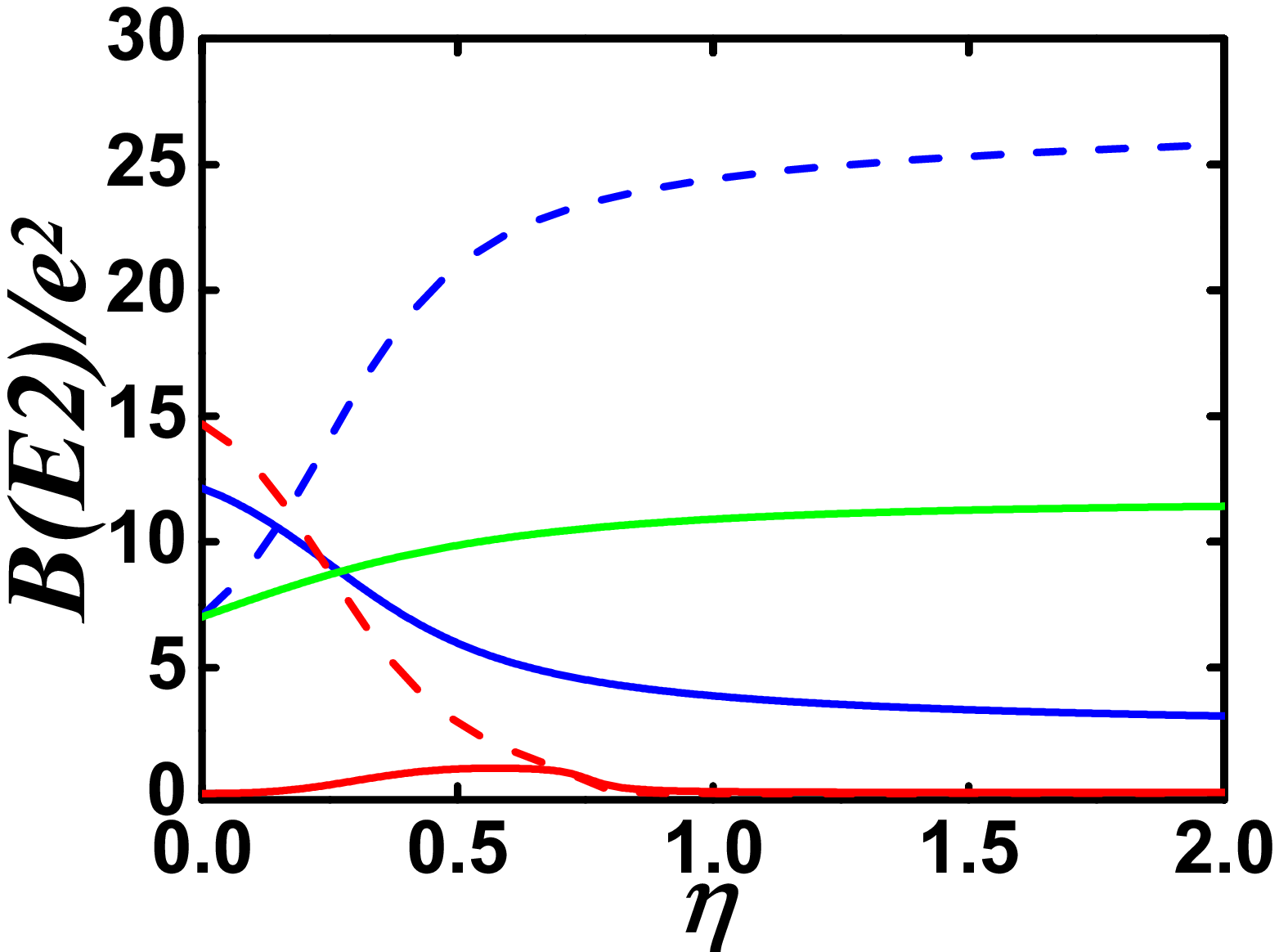}
\caption{Top: The evolutional behaviors of partial low-lying levels when $\eta$ changes from 0.0 to 2.0 for $N=7$ and $\alpha=1.235$ (other parameters are zero). Bottom: The evolutional behaviors of $B(E2; 2_{1}^{+}\rightarrow 0_{1}^{+})$ (green real line), $B(E2; 0_{2}^{+}\rightarrow 2_{1}^{+})$ (blue real line), $B(E2; 0_{2}^{+}\rightarrow 2_{2}^{+})$ (blue dashed line), $B(E2; 0_{3}^{+}\rightarrow 2_{1}^{+})$ (red real line), and $B(E2; 0_{3}^{+}\rightarrow 2_{2}^{+})$ (red dashed line) when $\eta$ changes from 0.0 to 2.0 for the same parameters as top.}
\end{figure}

\section{Hamiltonian in the SU3-IBM}

The Hamiltonian in the SU3-IBM has been discussed in Ref. \cite{Wang20,wang22,wang23,zhou23,Zhang22}. In previous paper, except for the $d$ boson number operator $\hat{n}_{d}$, only the SU(3) second-order and third-order Casimir operators $-\hat{C}_{2}[SU(3)]$, $\hat{C}_{3}[SU(3)]$ are considered to explain the normal states of $^{110}$Cd \cite{wang22}. When other SU(3) higher-order interactions are added, the Hamiltonian is as follows \cite{wang}
\begin{eqnarray}
\hat{H}&=&c\{\hat{n}_{d}+\eta[-\frac{\hat{C}_{2}[SU(3)]}{2N}+\alpha\frac{\hat{C}_{3}[SU(3)]}{2N^{2}}   \nonumber\\
&&+\beta\frac{\hat{C}_{2}^{2}[SU(3)]}{2N^{3}}+\gamma\frac{\Omega}{2N^{2}}+\delta\frac{\Lambda}{2N^{3}}]\},
\end{eqnarray}
where $c$, $\eta$, $\alpha$, $\beta$, $\gamma$, $\delta$ are fitting parameters. $\hat{Q}$ is the SU(3) quadrupole operator. $\Omega$ is $[\hat{L}\times \hat{Q} \times \hat{L}]^{(0)}$ and $\Lambda$ is $[(\hat{L}\times \hat{Q})^{(1)} \times (\hat{L} \times \hat{Q})^{(1)}]^{(0)}$. These two quantities result from the SU(3) mapping of the rigid triaxial rotor \cite{Isacker00,zhang14}. The $-\hat{C}_{2}[SU(3)]$, $\hat{C}_{3}[SU(3)]$ and $-\hat{C}_{2}^{2}[SU(3)]$ can make any SU(3) irreducible representation $(\lambda,\mu)$ become the ground state in the SU(3) limit. The study of SU3-IBM revealed that higher-order interactions are essential for describing the rotor rotation. At this point, many researchers have ignored it. This also makes IBM become very useful in understanding some of the collective behaviors.

The two SU(3) Casimir operators have relationships with the quadrupole second or third-order interactions in the SU(3) limit as follows:
\begin{equation}
	\hat{C}_{2}[\textrm{SU(3)}]=2\hat{Q}\cdot\hat{Q}+\frac{3}{4}\hat{L}\cdot\hat{L},
\end{equation}
\begin{equation}
	\hat{C}_{3}[\textrm{SU(3)}]=-\frac{4}{9}\sqrt{35}[\hat{Q}\times\hat{Q}\times\hat{Q}]^{(0)}-\frac{\sqrt{15}}{2}[\hat{L}\times\hat{Q}\times\hat{L}]^{(0)}.
\end{equation}
For a given SU(3) irreducible representation $(\lambda,\mu)$,the eigenvalues of the two Casimir operators under the group chain $U(6) \supset SU(3) \supset O(3)$ are expressed as
\begin{equation}
	\hat{C}_{2}[\textrm{SU(3)}]=\lambda^{2}+\mu^{2}+\mu\lambda+3\lambda+3\mu,
\end{equation}
\begin{equation}
	\hat{C}_{3}[\textrm{SU(3)}]=\frac{1}{9}(\lambda-\mu)(2\lambda+\mu+3)(\lambda+2\mu+3).
\end{equation}

\begin{figure}[tbh]
\includegraphics[scale=0.32]{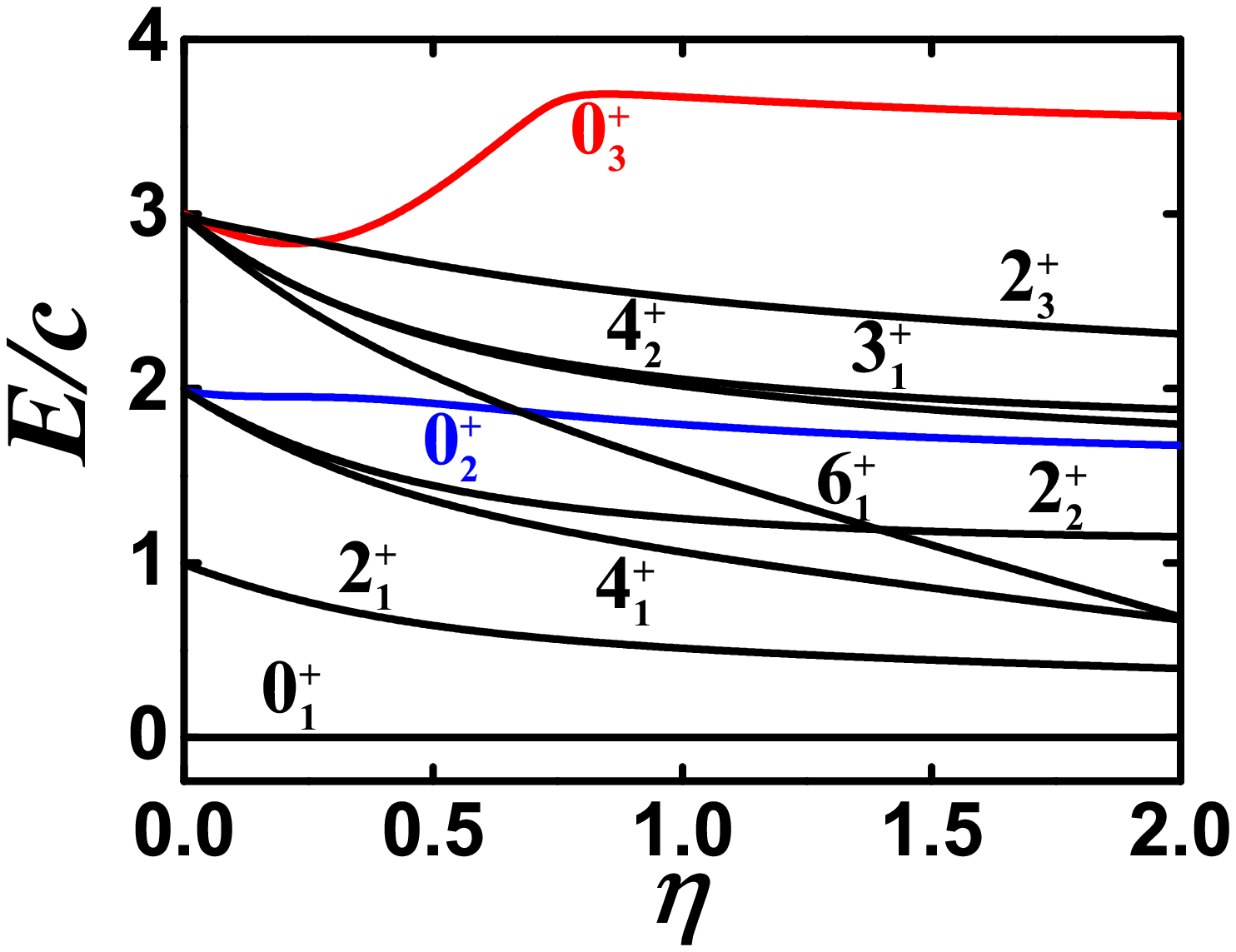}
\includegraphics[scale=0.32]{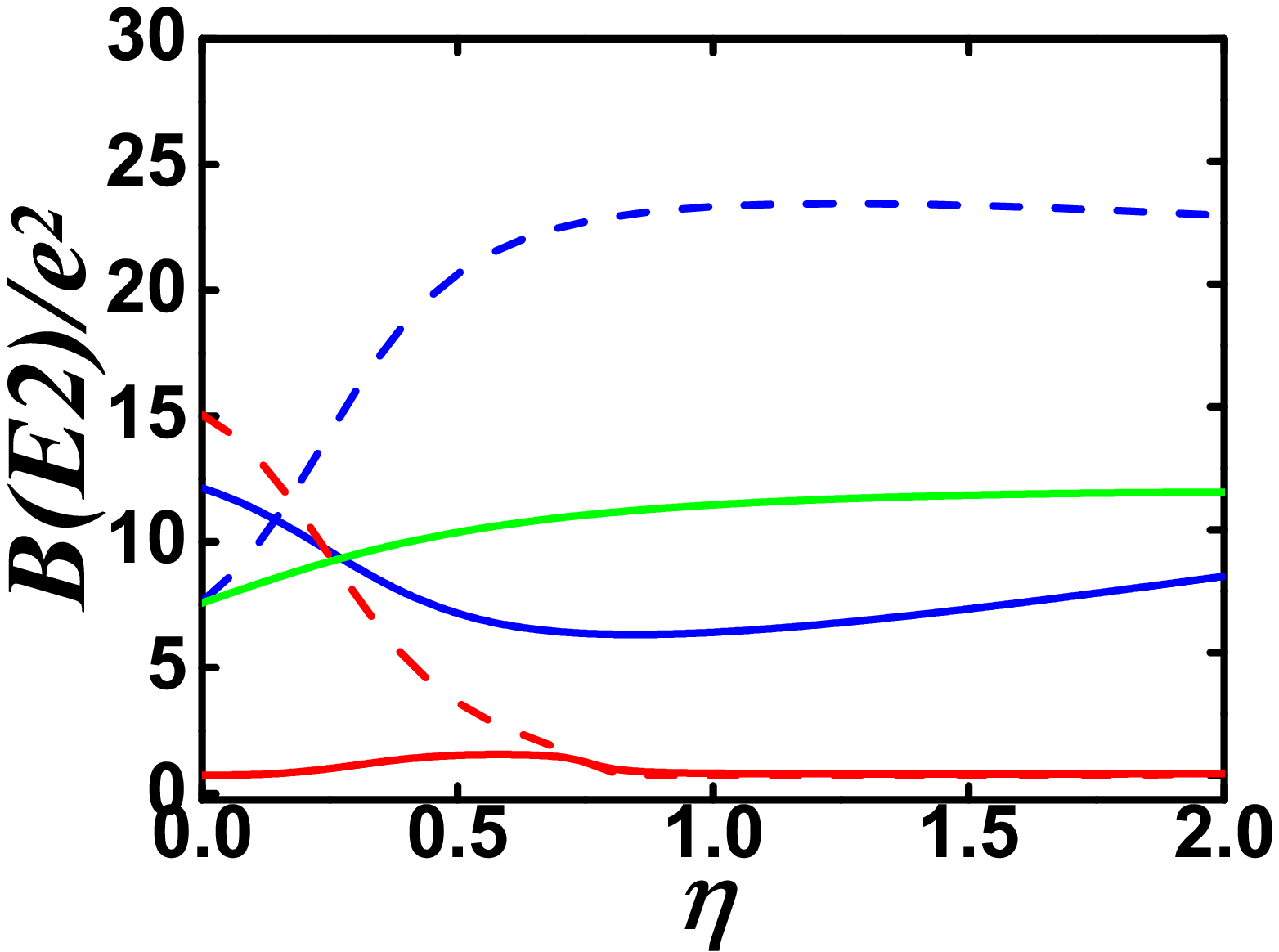}
\caption{Top: The evolutional behaviors of partial low-lying levels when $\eta$ changes from 0.0 to 2.0 for $N=7$, $\alpha=1.235$ and $\gamma=1.0$ (other parameters are zero). Bottom: The evolutional behaviors of $B(E2; 2_{1}^{+}\rightarrow 0_{1}^{+})$ (green real line), $B(E2; 0_{2}^{+}\rightarrow 2_{1}^{+})$ (blue real line), $B(E2; 0_{2}^{+}\rightarrow 2_{2}^{+})$ (blue dashed line), $B(E2; 0_{3}^{+}\rightarrow 2_{1}^{+})$ (red real line), and $B(E2; 0_{3}^{+}\rightarrow 2_{2}^{+})$ (red dashed line) when $\eta$ changes from 0.0 to 2.0 for the same parameters as top.}
\end{figure}

\begin{figure}[tbh]
\includegraphics[scale=0.32]{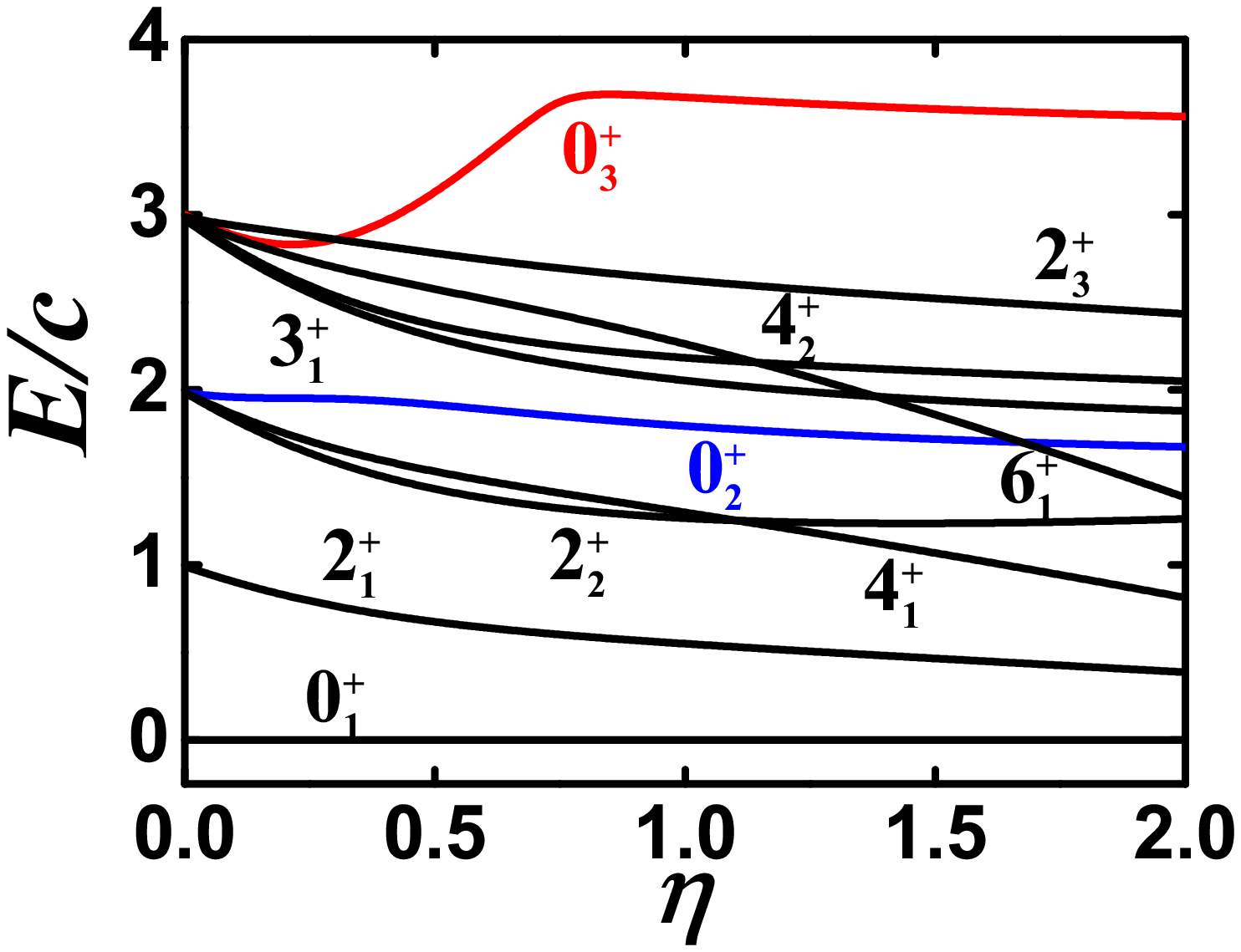}
\includegraphics[scale=0.32]{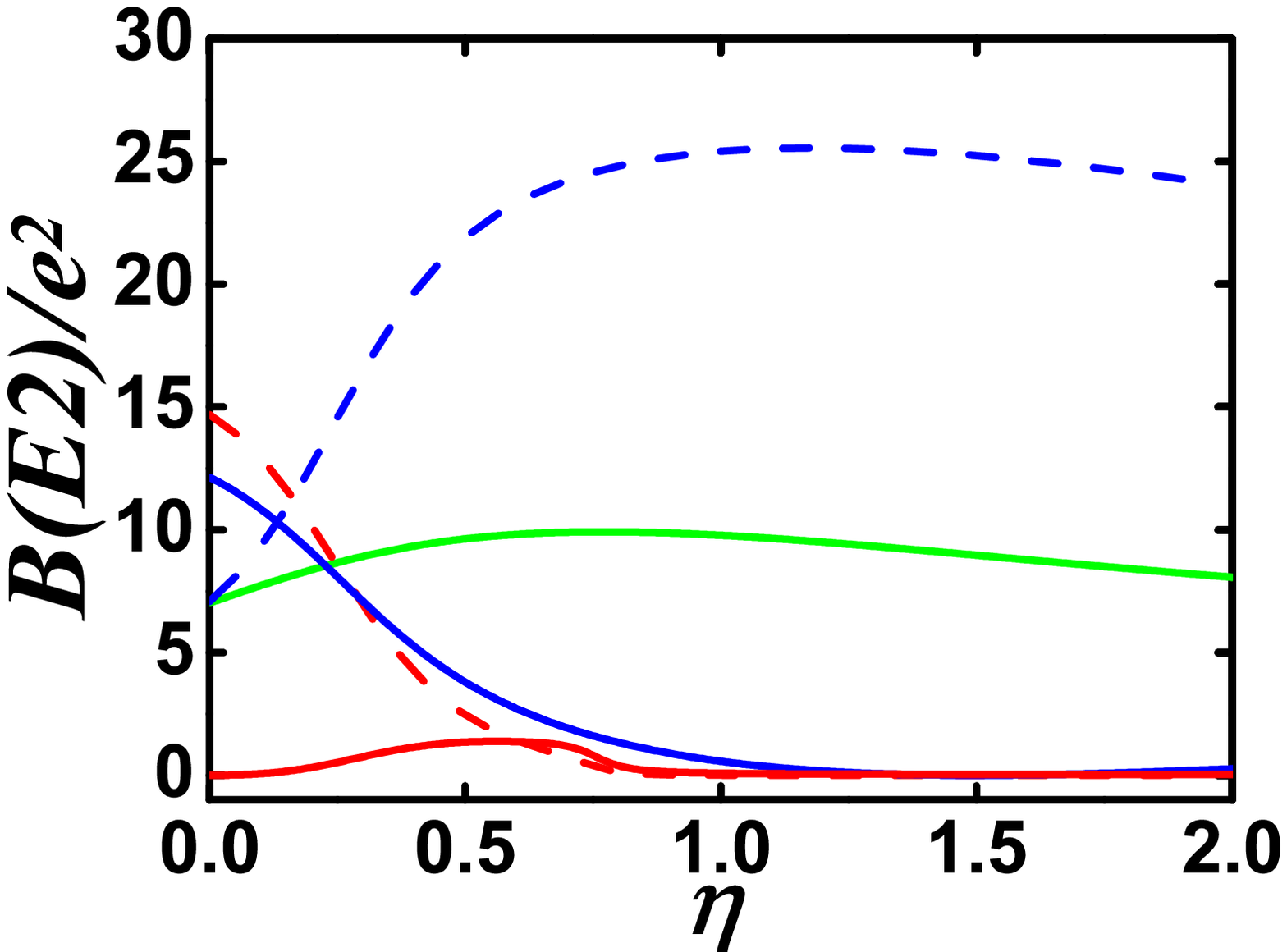}
\caption{Top: The evolutional behaviors of partial low-lying levels when $\eta$ changes from 0.0 to 2.0 for $N=7$, $\alpha=1.235$ and $\gamma=-2.0$ (other parameters are zero). Bottom: The evolutional behaviors of $B(E2; 2_{1}^{+}\rightarrow 0_{1}^{+})$ (green real line), $B(E2; 0_{2}^{+}\rightarrow 2_{1}^{+})$ (blue real line), $B(E2; 0_{2}^{+}\rightarrow 2_{2}^{+})$ (blue dashed line), $B(E2; 0_{3}^{+}\rightarrow 2_{1}^{+})$ (red real line), and $B(E2; 0_{3}^{+}\rightarrow 2_{2}^{+})$ (red dashed line) when $\eta$ changes from 0.0 to 2.0 for the same parameters as top.}
\end{figure}

\begin{figure}[tbh]
\includegraphics[scale=0.32]{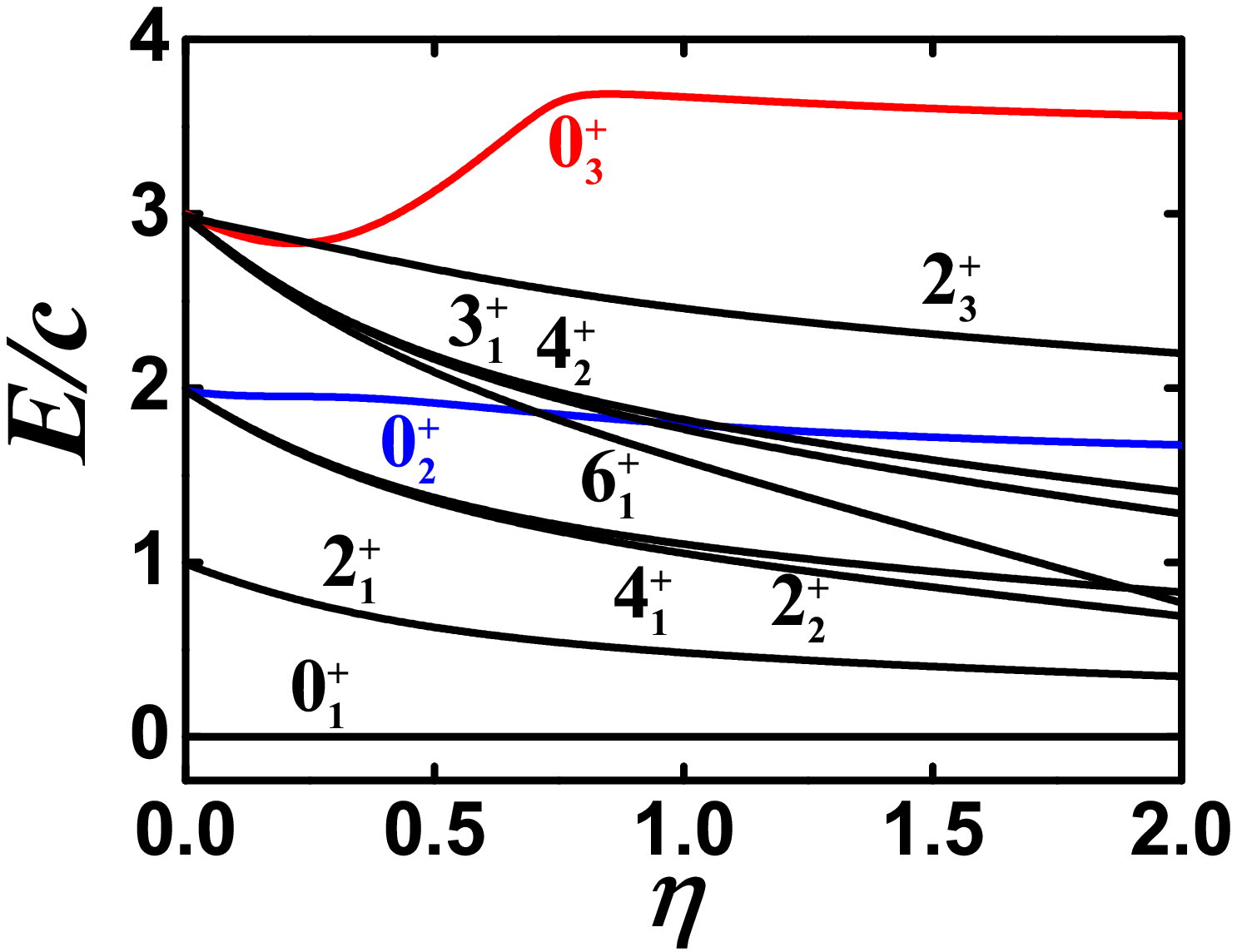}
\includegraphics[scale=0.32]{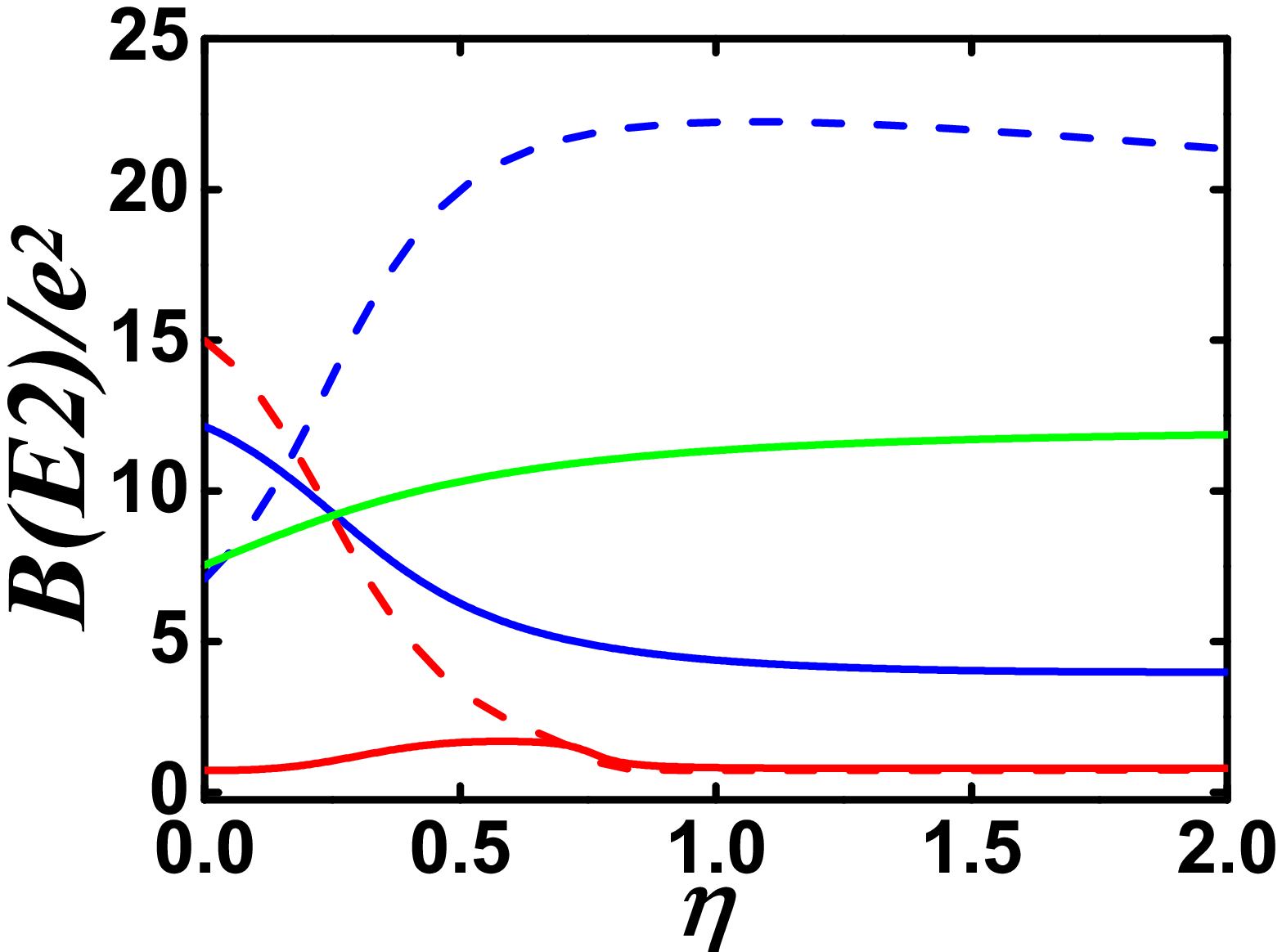}
\caption{Top: The evolutional behaviors of partial low-lying levels when $\eta$ changes from 0.0 to 2.0 for $N=7$, $\alpha=1.235$ and $\delta=2.0$ (other parameters are zero). Bottom: The evolutional behaviors of $B(E2; 2_{1}^{+}\rightarrow 0_{1}^{+})$ (green real line), $B(E2; 0_{2}^{+}\rightarrow 2_{1}^{+})$ (blue real line), $B(E2; 0_{2}^{+}\rightarrow 2_{2}^{+})$ (blue dashed line), $B(E2; 0_{3}^{+}\rightarrow 2_{1}^{+})$ (red real line), and $B(E2; 0_{3}^{+}\rightarrow 2_{2}^{+})$ (red dashed line) when $\eta$ changes from 0.0 to 2.0 for the same parameters as top.}
\end{figure}

\begin{figure}[tbh]
\includegraphics[scale=0.32]{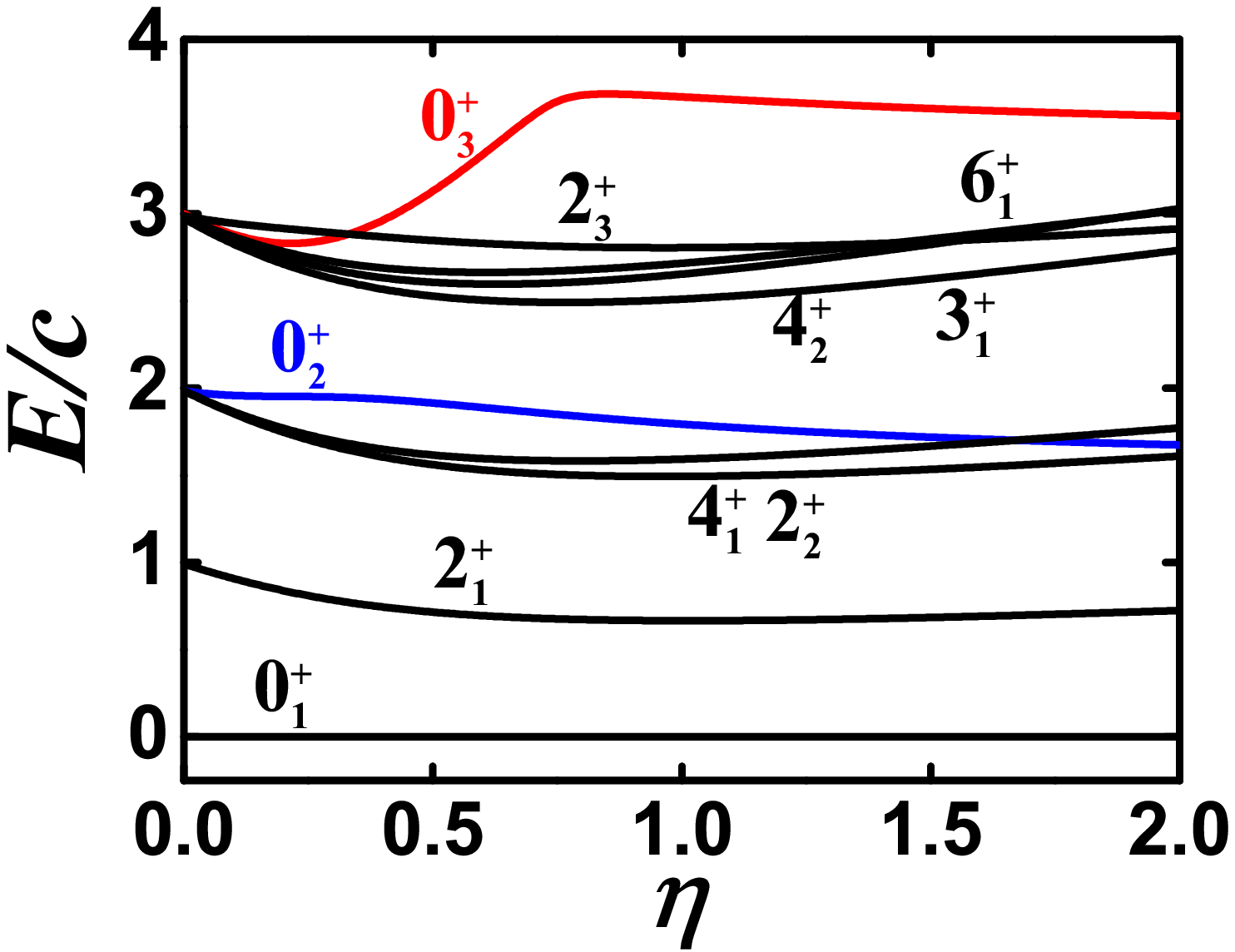}
\includegraphics[scale=0.32]{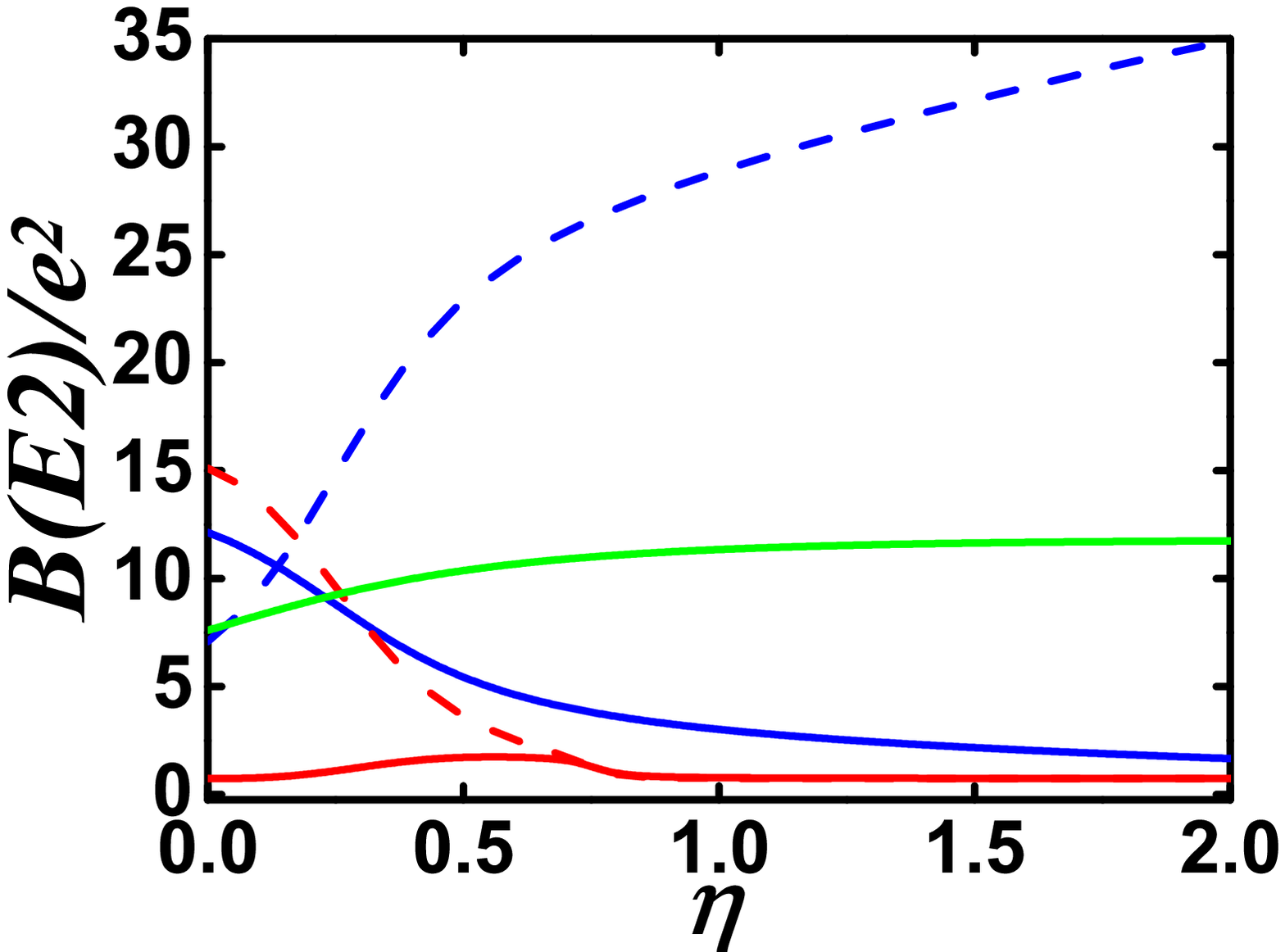}
\caption{Top: The evolutional behaviors of partial low-lying levels when $\eta$ changes from 0.0 to 2.0 for $N=7$, $\alpha=1.235$ and $\delta=-4.0$ (other parameters are zero). Bottom: The evolutional behaviors of $B(E2; 2_{1}^{+}\rightarrow 0_{1}^{+})$ (green real line), $B(E2; 0_{2}^{+}\rightarrow 2_{1}^{+})$ (blue real line), $B(E2; 0_{2}^{+}\rightarrow 2_{2}^{+})$ (blue dashed line), $B(E2; 0_{3}^{+}\rightarrow 2_{1}^{+})$ (red real line), and $B(E2; 0_{3}^{+}\rightarrow 2_{2}^{+})$ (red dashed line) when $\eta$ changes from 0.0 to 2.0 for the same parameters as top.}
\end{figure}

\begin{figure}[tbh]
\includegraphics[scale=0.32]{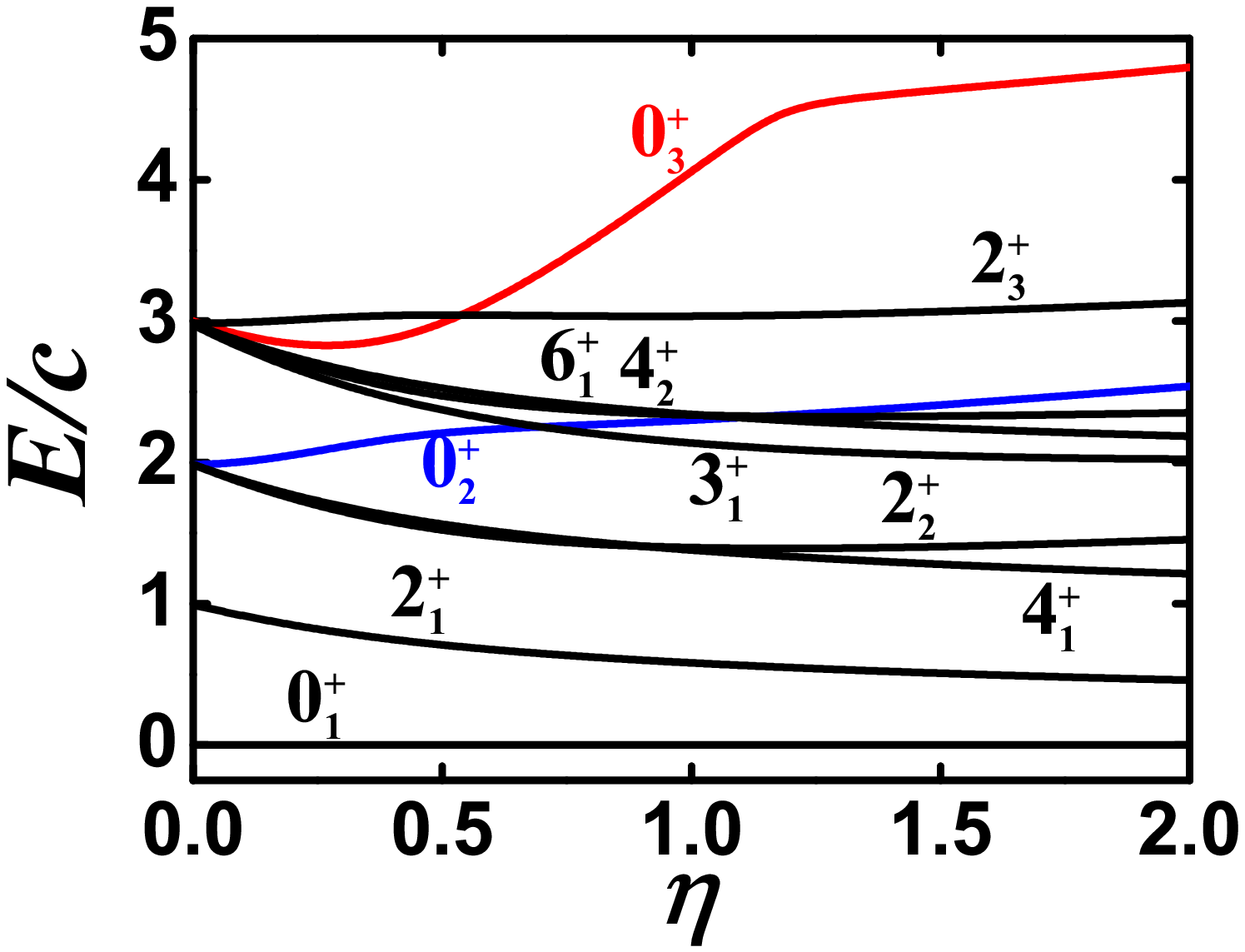}
\includegraphics[scale=0.32]{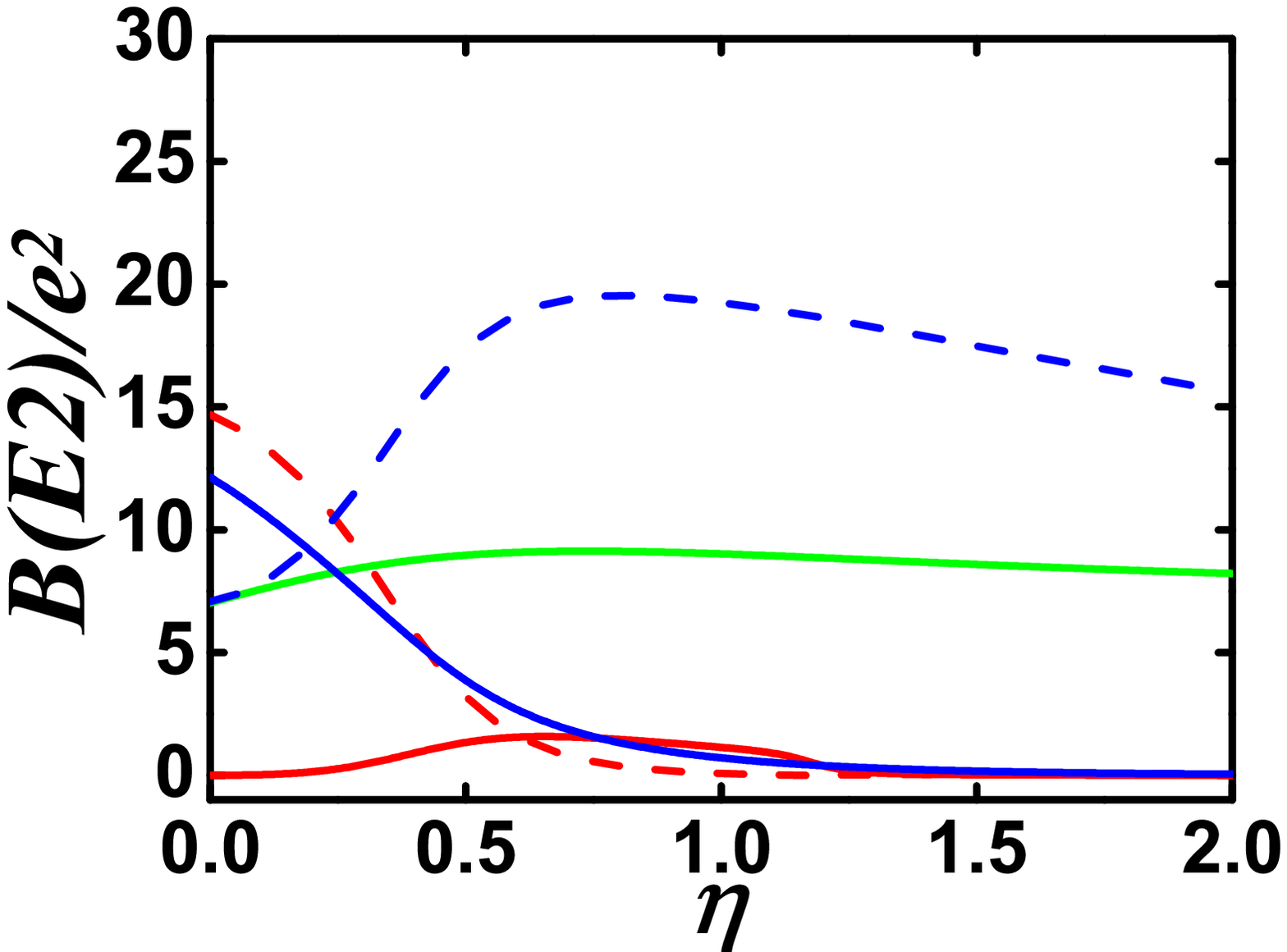}
\caption{Top: The evolutional behaviors of partial low-lying levels when $\eta$ changes from 0.0 to 2.0 for $N=7$, $\alpha=1.235$ and $\beta=0.03$ (other parameters are zero). Bottom: The evolutional behaviors of $B(E2; 2_{1}^{+}\rightarrow 0_{1}^{+})$ (green real line), $B(E2; 0_{2}^{+}\rightarrow 2_{1}^{+})$ (blue real line), $B(E2; 0_{2}^{+}\rightarrow 2_{2}^{+})$ (blue dashed line), $B(E2; 0_{3}^{+}\rightarrow 2_{1}^{+})$ (red real line), and $B(E2; 0_{3}^{+}\rightarrow 2_{2}^{+})$ (red dashed line) when $\eta$ changes from 0.0 to 2.0 for the same parameters as top.}
\end{figure}

In previous paper, only the $-\hat{C}_{2}[SU(3)]$ and $\hat{C}_{3}[SU(3)]$ were considered. $-\hat{C}_{2}[SU(3)]$ describes the prolate shape which has been known for many years. $\hat{C}_{3}(SU(3))$ presents the oblate shape, which was only recently enough to be taken seriously from the work \cite{Fortunato11}. In Ref. \cite{Zhang12}, the prolate-oblate shape phase transition from $-\hat{C}_{2}(SU(3))$ to $\hat{C}_{3}[SU(3)]$ was first discussed. Recently, when $\hat{n}_{d}$ is introduced, the asymmetric shape transition was studied in the Hf-Hg region \cite{wang23}. This result provides a solid experimental basis for the validity of SU3-IBM.

In the SU(3) description of the rigid triaxial rotor, the other SU(3) higher-order interactions are requisite \cite{Isacker00,zhang14}. Recently it was also found that explaining the B(E2) anomaly needs the introduction of other SU(3) higher-order interactions \cite{Wang20,Zhang22,Zhang24}.

\begin{figure}[tbh]
\includegraphics[scale=0.32]{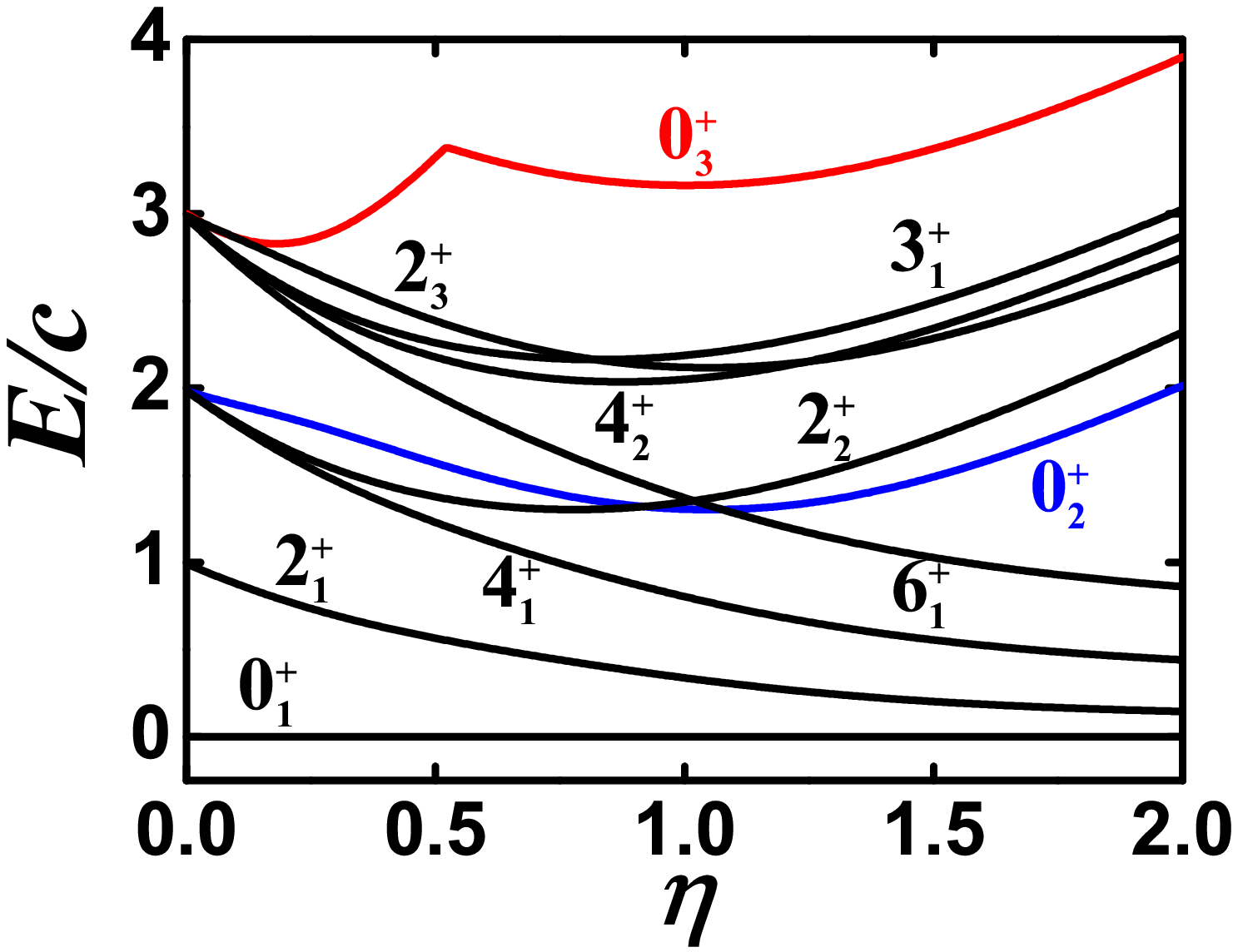}
\includegraphics[scale=0.32]{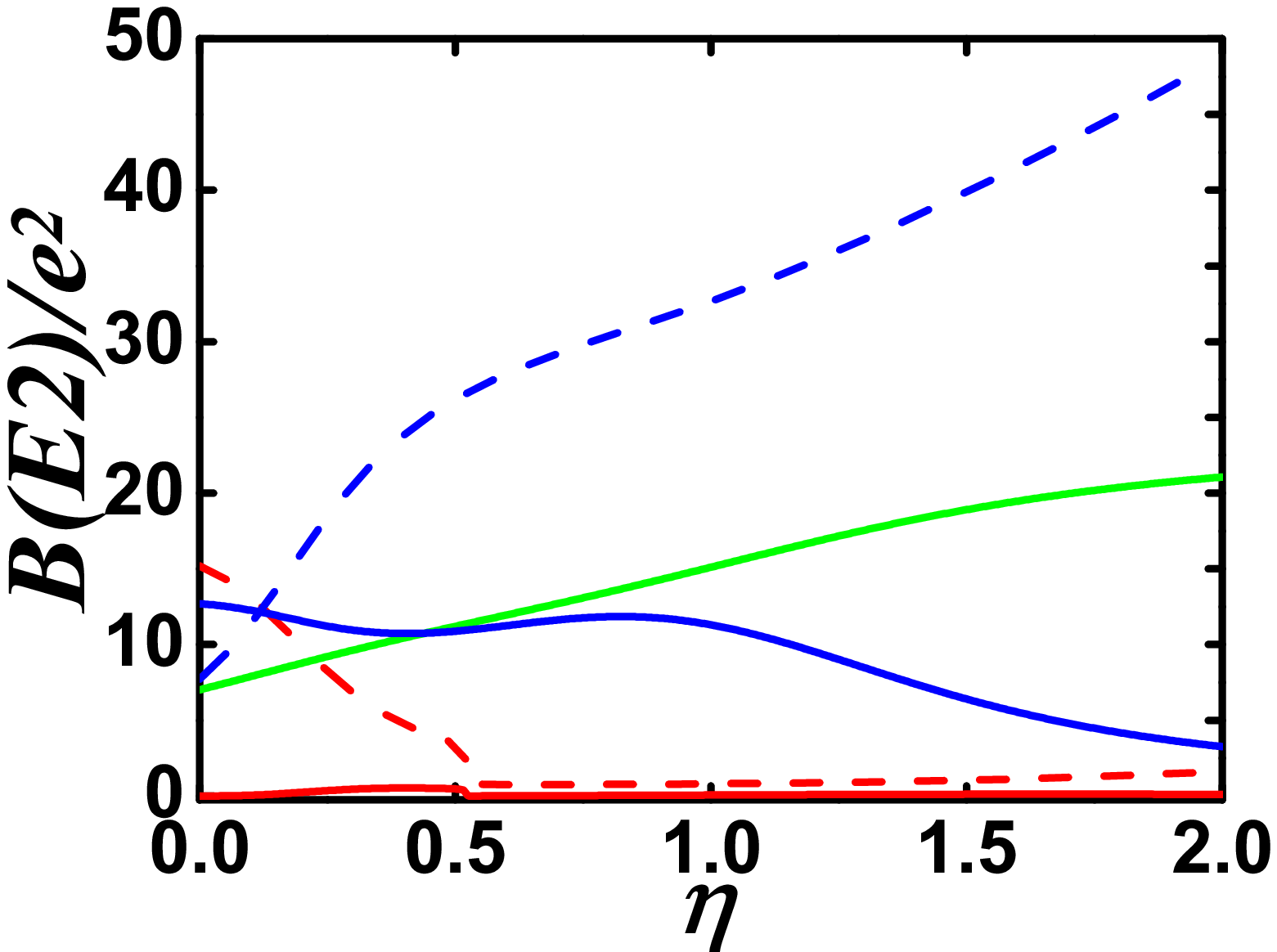}
\caption{Top: The evolutional behaviors of partial low-lying levels when $\eta$ changes from 0.0 to 2.0 for $N=7$, $\alpha=1.235$ and $\beta=-0.03$ (other parameters are zero). Bottom: The evolutional behaviors of $B(E2; 2_{1}^{+}\rightarrow 0_{1}^{+})$ (green real line), $B(E2; 0_{2}^{+}\rightarrow 2_{1}^{+})$ (blue real line), $B(E2; 0_{2}^{+}\rightarrow 2_{2}^{+})$ (blue dashed line), $B(E2; 0_{3}^{+}\rightarrow 2_{1}^{+})$ (red real line), and $B(E2; 0_{3}^{+}\rightarrow 2_{2}^{+})$ (red dashed line) when $\eta$ changes from 0.0 to 2.0 for the same parameters as top.}
\end{figure}

From $-\hat{C}_{2}[SU(3)]$ to $\hat{C}_{3}[SU(3)]$, there is a degenerate point for the SU(3) irreducible representation $(\lambda,\mu)$ satisfying the condition $\lambda+2\mu=2N$, which is also the shape phase transition point from the prolate shape to the oblate shape. $N$ is the boson number. The degenerate point is at $\alpha=\frac{3N}{2N+3}$. For the transitional region from the U(5) limit to the SU(3) degenerate point, degeneracy between the $4_{1}^{+}$ state and the $2_{2}^{+}$ state can be found \cite{wang22}, which is an typical feature of the $\gamma$-softness. Some other higher levels can be also degenerate. However, the reason for this degeneracy is still unknown. The key finding is that the middle positions of the transition region can show the spherical-like spectra for the explanation of the normal states of the Cd nuclei.

\begin{figure}[tbh]
\includegraphics[scale=0.32]{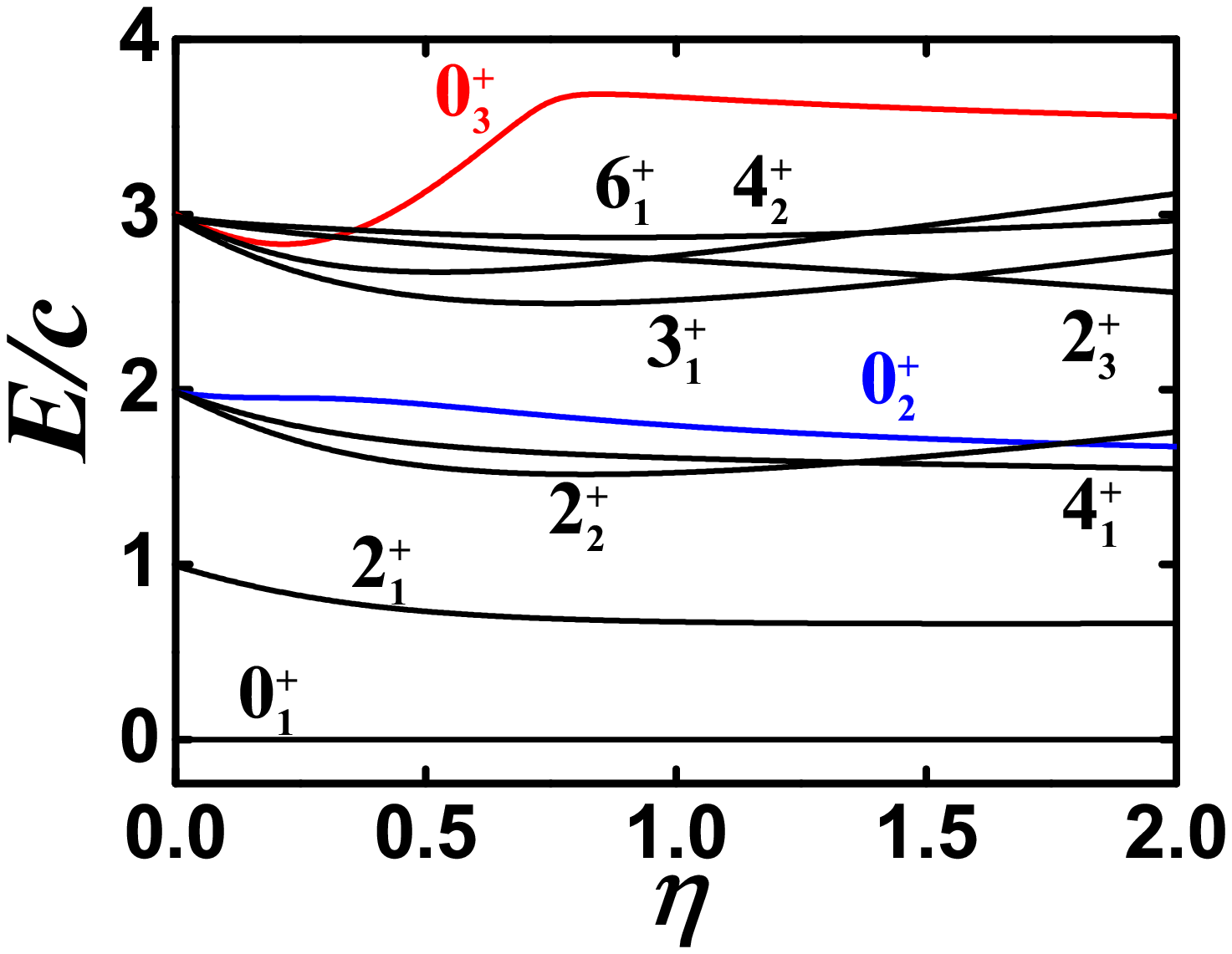}
\includegraphics[scale=0.32]{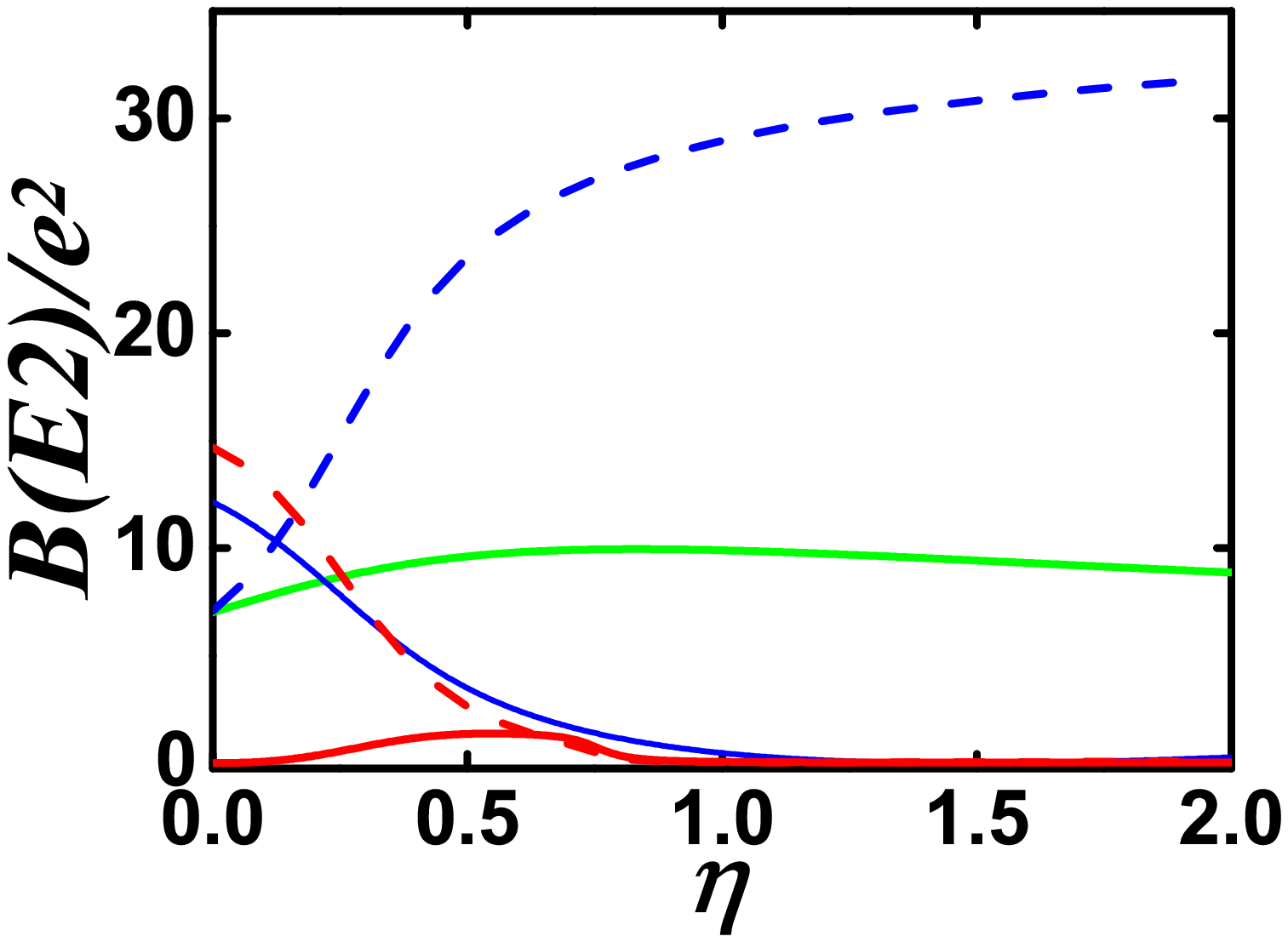}
\caption{Top: The evolutional behaviors of partial low-lying levels when $\eta$ changes from 0.0 to 2.0 for $N=7$, $\alpha=1.235$, $\beta=0$, $\gamma=-2.0$ and $\delta=-4.0$. Bottom: The evolutional behaviors of $B(E2; 2_{1}^{+}\rightarrow 0_{1}^{+})$ (green real line), $B(E2; 0_{2}^{+}\rightarrow 2_{1}^{+})$ (blue real line), $B(E2; 0_{2}^{+}\rightarrow 2_{2}^{+})$ (blue dashed line), $B(E2; 0_{3}^{+}\rightarrow 2_{1}^{+})$ (red real line), and $B(E2; 0_{3}^{+}\rightarrow 2_{2}^{+})$ (red dashed line) when $\eta$ changes from 0.0 to 2.0 for the same parameters as top.}
\end{figure}

Adding the $\hat{C}_{3}(SU(3))$ interaction has already described the spherical-like mode, but other SU(3) higher-order interactions are necessary when more details and configuration mixing is considered. Thus in this paper, the influences of other SU(3) higher-order interactions are investigated at a perturbation level. These are also important for understanding other nuclei with spherical nucleus puzzle, such as Te and Pd nuclei \cite{Wood18}.

\begin{figure}[tbh]
\includegraphics[scale=0.32]{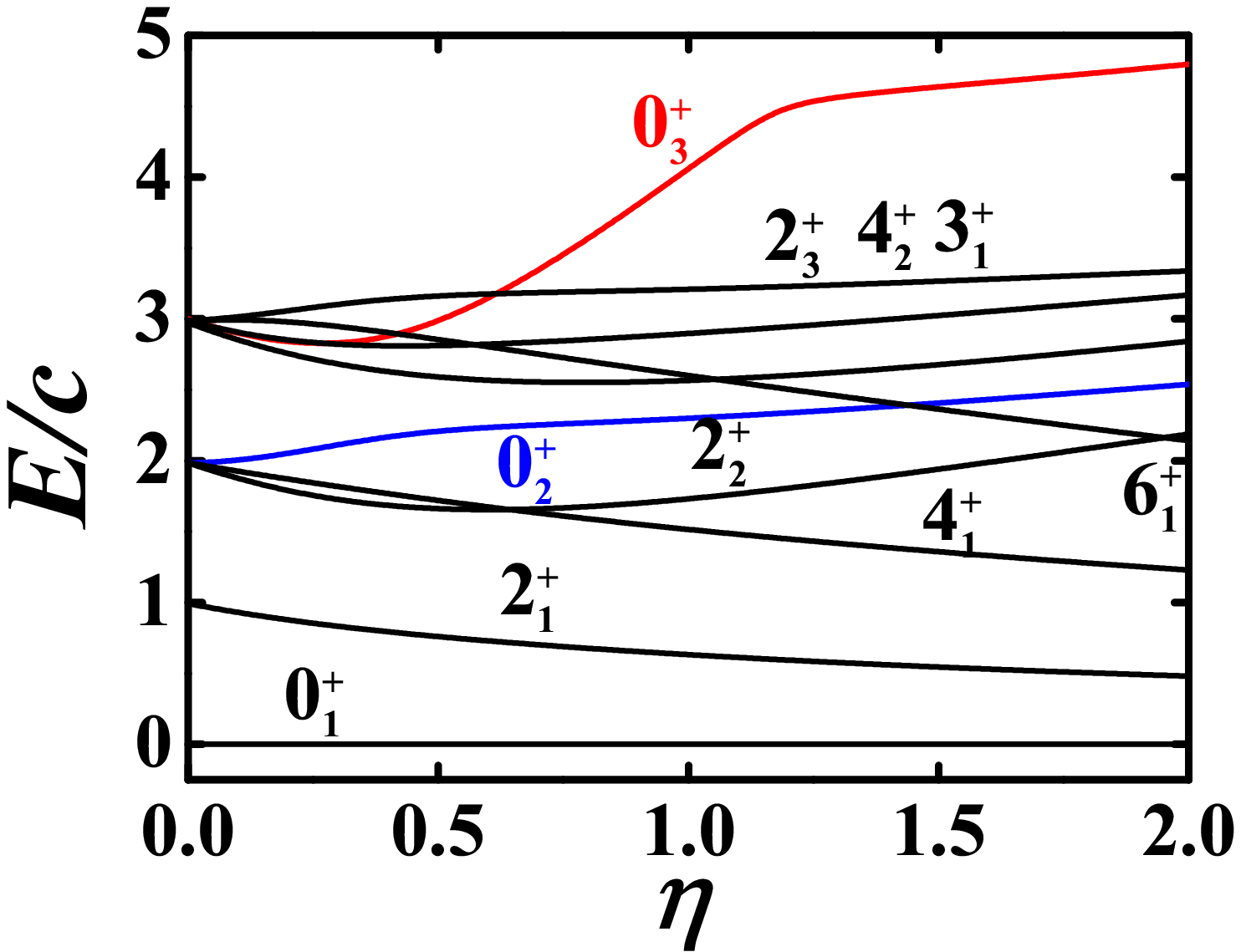}
\includegraphics[scale=0.32]{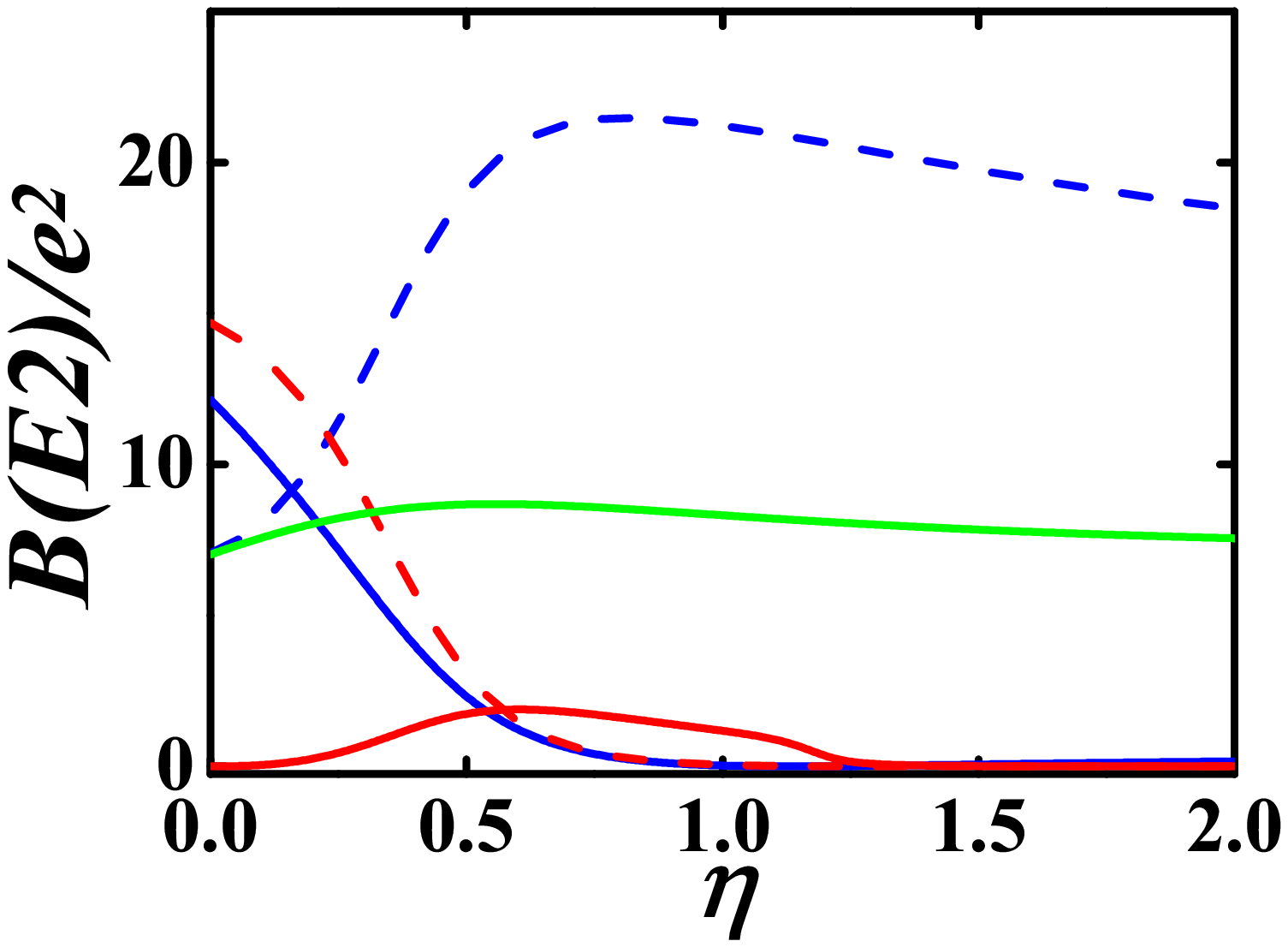}
\caption{Top: The evolutional behaviors of partial low-lying levels when $\eta$ changes from 0.0 to 2.0 for $N=7$, $\alpha=1.235$, $\beta=0.03$, $\gamma=-2.0$ and $\delta=-4.0$. Bottom: The evolutional behaviors of $B(E2; 2_{1}^{+}\rightarrow 0_{1}^{+})$ (green real line), $B(E2; 0_{2}^{+}\rightarrow 2_{1}^{+})$ (blue real line), $B(E2; 0_{2}^{+}\rightarrow 2_{2}^{+})$ (blue dashed line), $B(E2; 0_{3}^{+}\rightarrow 2_{1}^{+})$ (red real line), and $B(E2; 0_{3}^{+}\rightarrow 2_{2}^{+})$ (red dashed line) when $\eta$ changes from 0.0 to 2.0 for the same parameters as top.}
\end{figure}

For better understanding various $\gamma$-softness, the $B(E2)$ values are also necessary. How to distinguish the different $\gamma$-softness becomes more important in the investigation of the realistic nuclei properties. The $E2$ operator is defined as
\begin{equation}
\hat{T}(E2)=e\hat{Q},
\end{equation}
where $e$ is the boson effective charge. The evolution of the $B(E2; 2_{1}^{+}\rightarrow 0_{1}^{+})$, $B(E2; 0_{2}^{+}\rightarrow 2_{1}^{+})$, $B(E2; 0_{2}^{+}\rightarrow 2_{2}^{+})$, $B(E2; 0_{3}^{+}\rightarrow 2_{1}^{+})$, and $B(E2; 0_{3}^{+}\rightarrow 2_{2}^{+})$ values are discussed for various parameters.

\begin{figure}[tbh]
\includegraphics[scale=0.32]{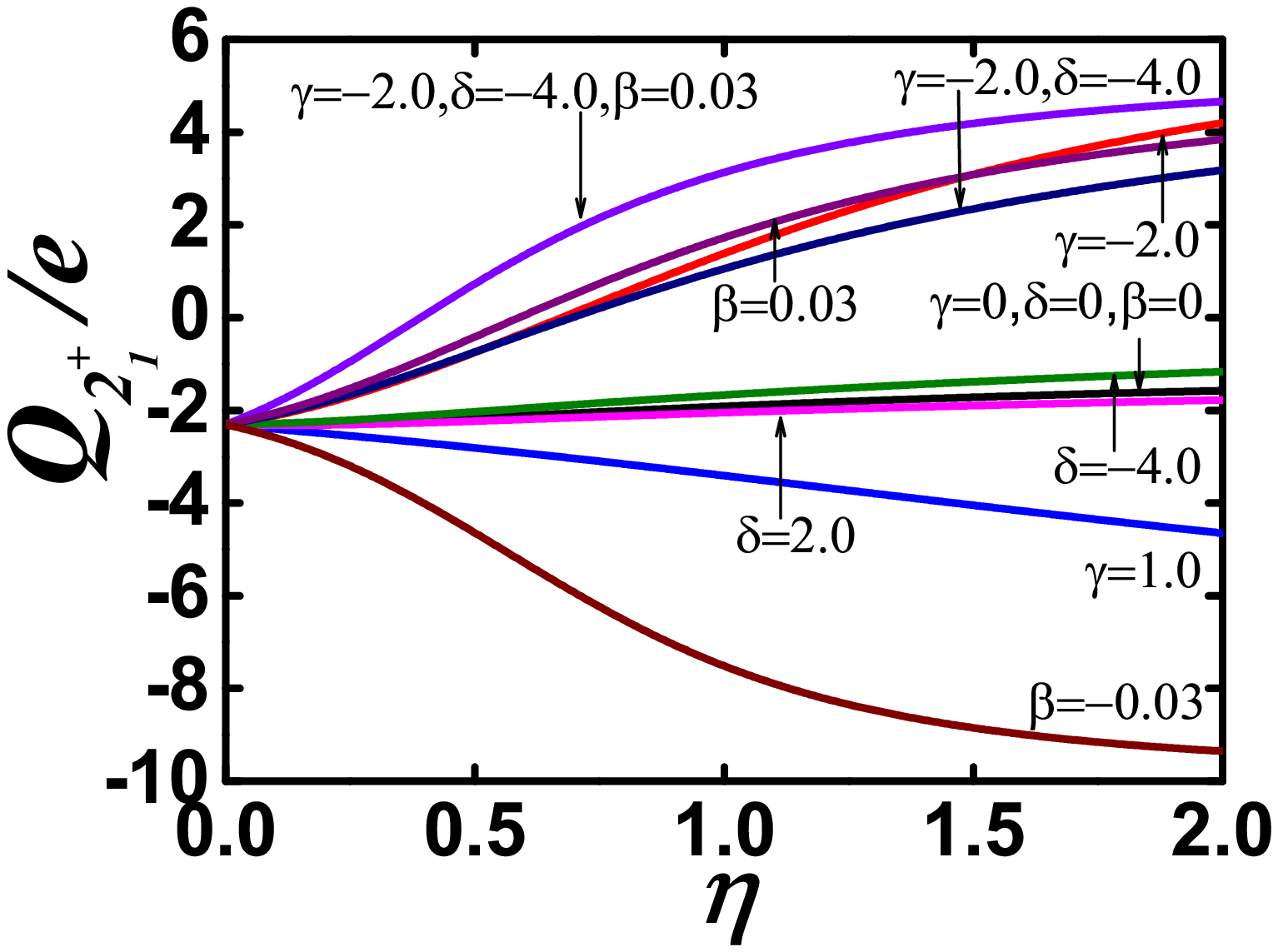}
\caption{Evolutional behaviors of $Q_{2_{1}^{+}}$ values whith $\eta$ when $\alpha=1.235$ and different other parameters are chosen.}
\end{figure}

\section{Influences of the SU(3) higher-order interactions}

When other SU(3) higher-order interactions are not considered, this case has been discussed in \cite{wang22}, which is shown in Fig. 4. It should be noticed that the Hamiltonian here is somewhat different from the one in \cite{wang22}, which makes the evolutional behaviors more explicit. The boson number is also chosen as 7. Obviously, the degeneracy between the $4_{1}^{+}$ and $2_{2}^{+}$ states can be found. Similar degenerate phenomenon can be also seen in the $6_{1}^{+}$, $4_{2}^{+}$, $3_{1}^{+}$ states. The $0_{2}^{+}$ state is between the two groups of degenerate levels. The $0_{3}^{+}$ state moves up obviously, which can not be degenerate with the $6_{1}^{+}$, $4_{2}^{+}$, $3_{1}^{+}$, $2_{3}^{+}$ states any more. It should be noticed that the $0_{2}^{+}$ state can not be higher than the $6_{1}^{+}$, $4_{2}^{+}$, $3_{1}^{+}$ states (This conclusion can be obtained by numerical calculation). These level features are very different from the ones in the O(6) symmetry, see Ref. \cite{wang}.

Although the low-lying levels have agreed well between the theoretical fitting results and the experimental data, the fitting effect of the B(E2) values among the levels is still insufficient in Ref. \cite{wang22}. Especially the B(E2) values between the $0_{2}^{+}$ state and the $2_{1}^{+}$ state can be nearly zero in $^{112-116}$Cd. The $0_{2}^{+}$ state seems to be nearly degenerate with the $4_{1}^{+}$ and $2_{2}^{+}$ states, which leads to the misunderstanding that the three states constitute the two-phonon triplet states. However the nearly zero B(E2) value between the $0_{2}^{+}$ state and the $2_{1}^{+}$ state show that phonon vibration model is simply impossible. This conclusion may be difficult for many researchers to accept it in the field of nuclear structure. In Fig. 4 (bottom graph), it is obvious that the B(E2) value between the $0_{2}^{+}$ state and the $2_{1}^{+}$ state is much smaller than the one between the $2_{1}^{+}$ state and the $0_{1}^{+}$ state. However it can not reach nearly zero. For further reducing the B(E2) value, other SU(3) higher-order interactions are needed. Moreover, in Fig. 4, the $0_{2}^{+}$ state is closer to the $6_{1}^{+}$, $4_{2}^{+}$, $3_{1}^{+}$, $2_{3}^{+}$ states. If it can bring the $0_{2}^{+}$ state closer to the $4_{1}^{+}$ and $2_{2}^{+}$  states, the introduction of the higher-order interactions seems more plausible.

Here, we focus on the position $\eta=1.0$ which shows a typical spherical-like spectra. For $\alpha=\frac{3N}{2N+3}$, this position is just the critical point between the prolate shape and the rigid triaxial shape \cite{wang23}. When other SU(3) higher order interactions are introduced at a perturbation level, the quadrupole deformation will be changed a little, more prolate or more oblate.

First, the O(3) scalar shift operator $\Omega=[\hat{L} \times \hat{Q} \times \hat{L}]^{(0)}$ is added to the model calculation which are shown in Fig. 5 and Fig. 6. The coefficient of this SU(3) third-order interaction in Fig. 5 is 1.0 while it is -2.0 in Fig. 6. Obviously negative coefficient can greatly reduce the B(E2) value between the $0_{2}^{+}$ state and the $2_{1}^{+}$ state. The degeneracy of the energy levels is broken, but it varies little.

Second, the SU(3) forth-order interaction $[(\hat{L} \times \hat{Q})^{(1)} \times (\hat{L} \times \hat{Q})^{(1)}]^{(0)}$ is added to the model calculation which are shown in Fig. 7 and Fig. 8. The coefficient is 2.0 in Fig. 7 while it is -4.0 in Fig. 8. The negative coefficient can reduce the B(E2) value between the $0_{2}^{+}$ state and the $2_{1}^{+}$ state a little. However the levels are more like the spectra in the U(5) limit except for the $0_{3}^{+}$ state. This is an important result.

Third, the $\hat{C}_{2}^{2}[SU(3)]$ is added to the model calculation which are shown in Fig. 9 and Fig. 10. The coefficient is 0.03 in Fig. 9 while it is -0.03 in Fig. 10. The positive coefficient can reduce the B(E2) value between the $0_{2}^{+}$ state and the $2_{1}^{+}$ state. The degeneracy of the energy levels is broken, but it varies little also.

Now the $\Omega=[\hat{L} \times \hat{Q} \times \hat{L}]^{(0)}$ and $[(\hat{L} \times \hat{Q})^{(1)} \times (\hat{L} \times \hat{Q})^{(1)}]^{(0)}$ are both added to the model calculation which is shown in Fig. 11. The coefficients of the two are -2.0 and -4.0. The energy spectra are very similar to the phonon excitation of a spherical nucleus except for the $0_{3}^{+}$ state, and the B(E2) value between the $0_{2}^{+}$ state and the $2_{1}^{+}$ state is very small. The spherical-like spectra really exist, and it is exactly what we want to find. In previous theories, it is hard to believe that this collective excitation is indeed possible. This also shows that the SU3-IBM is completely self-consistent.

For completeness, the three higher-order interactions are all added to the model calculation which is shown in Fig. 12 with the coefficients -2.0, -4.0, 0.03. The B(E2) value between the $0_{2}^{+}$ state and the $2_{1}^{+}$ state is nearly zero. However, the B(E2) value between the $0_{3}^{+}$ state and the $2_{1}^{+}$ state becomes larger.

Last, the quadrupole moments of the $2_{1}^{+}$ state $Q_{2_{1}^{+}}$ are discussed when various parameters are chosen which is shown in Fig. 13. When other SU(3) higher-order interactions are not considered, the $Q_{2_{1}^{+}}$ value is about -2.0e, which means a prolate shape. The $Q_{2_{1}^{+}}$ value is insensitive to the fourth-order interaction $[(\hat{L} \times \hat{Q})^{(1)} \times (\hat{L} \times \hat{Q})^{(1)}]^{(0)}$ (parameter $\delta$) which can induce a prominent spherical-like spectra. Thus this interaction is necessary. The introduction of $\Omega=[\hat{L} \times \hat{Q} \times \hat{L}]^{(0)}$ and $\hat{C}_{2}^{2}[SU(3)]$ can make the B(E2) values of the $0_{2}^{+}$ state and the $2_{1}^{+}$ state very small, but they also let the $Q_{2_{1}^{+}}$ values shift to the positive side and the nucleus becomes an oblate shape. This conflicts with the experimental results. Thus these quantities may be needed, but the parameters of them are not too large. These findings provide strong constraints on the SU(3) higher-order interactions, which are very important for the fitting, as well as for configuration \textbf{mixing} calculations in future. In the latter fitting, it will be seen that $Q_{2_{1}^{+}}$ plays an important role. Our calculation can reproduce the unique evolution trend of the the quadrupole moments of the $2_{1}^{+}$ state $Q_{2_{1}^{+}}$ in $^{108-116}$Cd, which proves the validity of the conjecture on the spherical-like $\gamma$-soft mode.

\begin{figure}[tbh]
\includegraphics[scale=0.55]{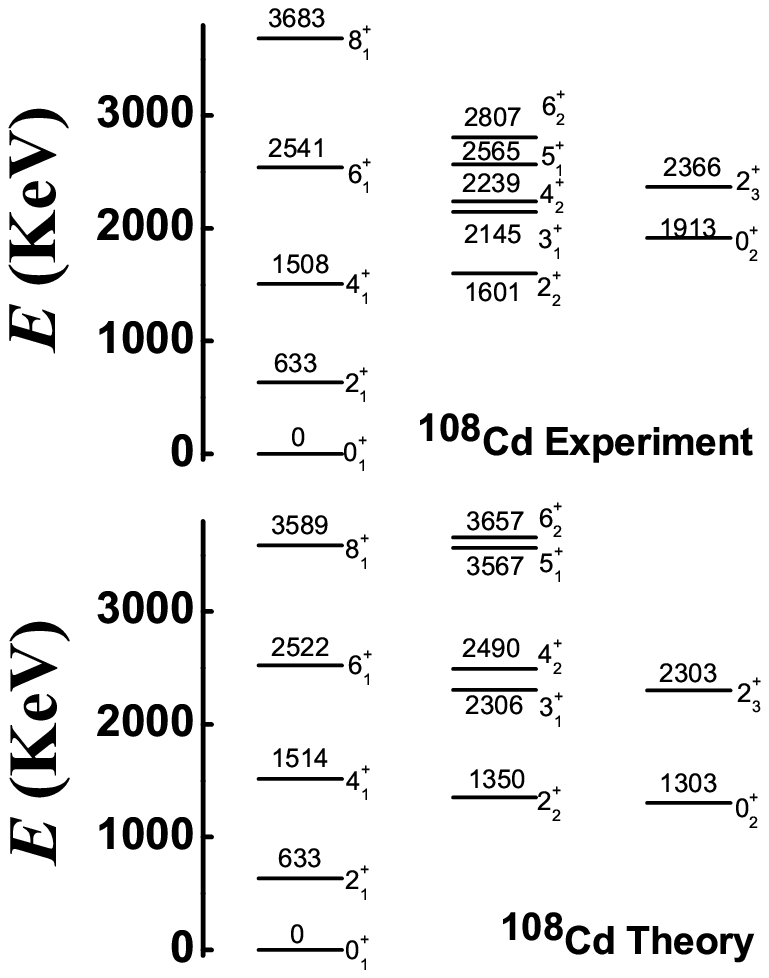}
\caption{Energy spectra of the normal states of $^{108}$Cd (Top: Experiment, Bottom: Theory).}
\end{figure}

\begin{table}[!ht]
    \centering
    \caption{Absolute B(E2) values in W.u. for $E2$ transitions from the low-lying normal states of $^{108}$Cd with effective charge $e=1.78$ (W.u.)$^{1/2}$. The last row is the $Q_{2_{1}^{+}}$ value in eb.}
    \begin{tabular}{llllll}
    \hline
        B(E2)~~~~ & Exp. & Theory~~~~~~ &B(E2)~~~~ & Exp. & Theory \\ \hline
        2$_{1}^{+}$$\rightarrow$0$_{1}^{+}$  &26.6(3)   & 26.6           & 4$_{1}^{+}$$\rightarrow$2$_{1}^{+}$ & 41(6) &31.86 \\
        2$_{2}^{+}$$\rightarrow$2$_{1}^{+}$  & 17(5)  &38.89 & 2$_{2}^{+}$$\rightarrow$0$_{1}^{+}$ & 1.8(3) & 1.02\\
        2$_{3}^{+}$$\rightarrow$2$_{2}^{+}$  &   & 4.04               & 2$_{3}^{+}$$\rightarrow$0$_{2}^{+}$ & &18.89     \\
        0$_{2}^{+}$$\rightarrow$2$_{1}^{+}$  &    &4.07                & 0$_{2}^{+}$$\rightarrow$2$_{2}^{+}$ & & 90.41    \\
        6$_{1}^{+}$$\rightarrow$4$_{1}^{+}$  &  &27.83                & 6$_{1}^{+}$$\rightarrow$4$_{2}^{+}$ & &0.31     \\
        2$_{3}^{+}$$\rightarrow$4$_{1}^{+}$  &  & 0.13               & 2$_{3}^{+}$$\rightarrow$2$_{1}^{+}$ & & 1.02    \\
        4$_{2}^{+}$$\rightarrow$4$_{1}^{+}$  &  & 14.79               & 4$_{2}^{+}$$\rightarrow$2$_{1}^{+}$ & &2.39     \\
        4$_{2}^{+}$$\rightarrow$2$_{2}^{+}$  &  &25.44                & 3$_{1}^{+}$$\rightarrow$4$_{1}^{+}$ & &18.83     \\
        3$_{1}^{+}$$\rightarrow$2$_{1}^{+}$  & & 0.41               & 3$_{1}^{+}$$\rightarrow$2$_{2}^{+}$ & & 20.61    \\
        0$_{3}^{+}$$\rightarrow$2$_{1}^{+}$  &  &0.026                & 0$_{3}^{+}$$\rightarrow$2$_{2}^{+}$ & & 0.2 \\  \hline
        $Q_{2^{+}_{1}}$  &-0.45 & -0.086\\ \hline
    \end{tabular}
    \label{dsafdsfdfd}
\end{table}

\begin{figure}[tbh]
\includegraphics[scale=0.55]{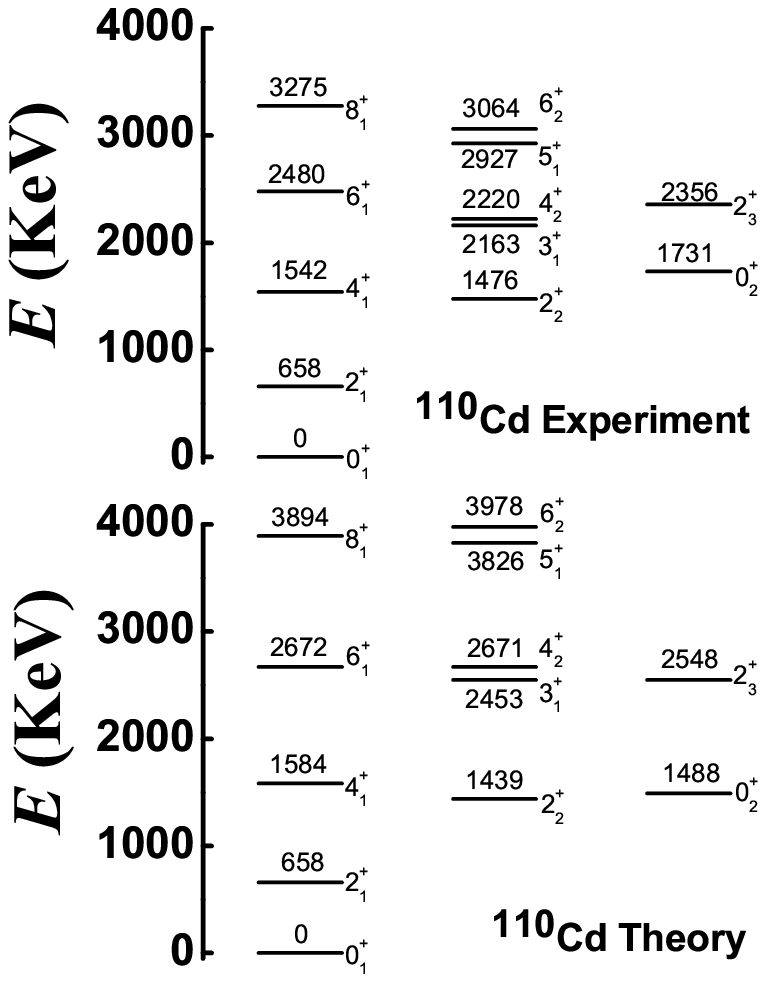}
\caption{Energy spectra of the normal states of $^{110}$Cd (Top: Experiment, Bottom: Theory).}
\end{figure}

\begin{table}[!ht]
    \centering
    \caption{Absolute B(E2) values in W.u. for $E2$ transitions from the low-lying normal states of $^{110}$Cd with effective charge $e=1.59$ (W.u.)$^{1/2}$. The last row is the $Q_{2_{1}^{+}}$ value in eb.}
    \begin{tabular}{llllll}
    \hline
        B(E2)~~~~ & Exp. &Theory~~~~~~ &B(E2)~~~~ & Exp. &Theory \\ \hline
        2$_{1}^{+}$$\rightarrow$0$_{1}^{+}$  &27.0(8)     & 27                   & 4$_{1}^{+}$$\rightarrow$2$_{1}^{+}$ & 42(9) &33.99 \\
        2$_{2}^{+}$$\rightarrow$2$_{1}^{+}$  & 30(5) & 39.85 & 2$_{2}^{+}$$\rightarrow$0$_{1}^{+}$ & 1.35(20)  &0.95 \\
        2$_{3}^{+}$$\rightarrow$2$_{2}^{+}$  & 0.7$_{-0.6}^{+0.5}$  & 3.55 & 2$_{3}^{+}$$\rightarrow$0$_{2}^{+}$ & 24.2(22) &20.17 \\
        0$_{2}^{+}$$\rightarrow$2$_{1}^{+}$  & $<$7.9  &  4.14            & 0$_{2}^{+}$$\rightarrow$2$_{2}^{+}$ &$<$1680$^{a}$ & 84.05 \\
        6$_{1}^{+}$$\rightarrow$4$_{1}^{+}$  & 40(30)  &35.512 & 6$_{1}^{+}$$\rightarrow$4$_{2}^{+}$ & $<$5$^{a}$ & 0.33\\
        2$_{3}^{+}$$\rightarrow$4$_{1}^{+}$  &$<$5$^{a}$  &0.3 & 2$_{3}^{+}$$\rightarrow$2$_{1}^{+}$ & 2.8$_{-1.0}^{+0.6}$  &0.71 \\
        4$_{2}^{+}$$\rightarrow$4$_{1}^{+}$  & 12$_{-6.0}^{+4.0}$ &16.13 & 4$_{2}^{+}$$\rightarrow$2$_{1}^{+}$ & 0.2$_{-0.09}^{+0.06}$ & 2.02\\
        4$_{2}^{+}$$\rightarrow$2$_{2}^{+}$  &32$_{-14}^{+10}$  & 26.59 & 3$_{1}^{+}$$\rightarrow$4$_{1}^{+}$ & 5.9$_{-4.6}^{+1.8}$& 18.67\\
        3$_{1}^{+}$$\rightarrow$2$_{1}^{+}$  & 1.1$_{-0.8}^{+1.3}$  & 0.47& 3$_{1}^{+}$$\rightarrow$2$_{2}^{+}$ & 38$_{-24}^{+8}$& 23.50\\
        0$_{3}^{+}$$\rightarrow$2$_{1}^{+}$  &  &  0.072                    & 0$_{3}^{+}$$\rightarrow$2$_{2}^{+}$ & & 0.13 \\\hline
        $Q_{2^{+}_{1}}$  & -0.40 & -0.076  \\ \hline
    \end{tabular}
    \label{dsafdsfdfd}
\end{table}

\begin{figure}[tbh]
\includegraphics[scale=0.55]{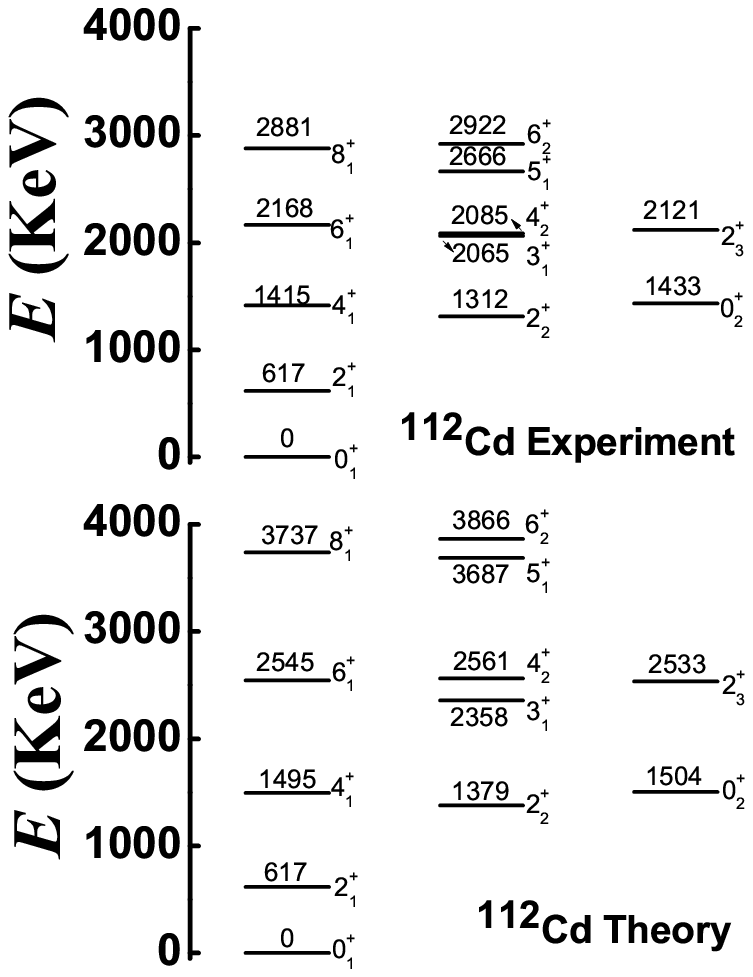}
\caption{Energy spectra of the normal states of $^{112}$Cd (Top: Experiment, Bottom: Theory).}
\end{figure}

\begin{table}[!ht]
    \centering
    \caption{Absolute B(E2) values in W.u. for $E2$ transitions from the low-lying normal states of $^{112}$Cd with effective charge $e=1.53$ (W.u.)$^{1/2}$. The last row is the $Q_{2_{1}^{+}}$ value in eb.}
    \begin{tabular}{llllll}
    \hline
        B(E2)~~~~ & Exp. &Theory~~~~~~ &B(E2)~~~~ & Exp. &Theory \\ \hline
        2$_{1}^{+}$$\rightarrow$0$_{1}^{+}$  &30.31(19)     & 30.31                   & 4$_{1}^{+}$$\rightarrow$2$_{1}^{+}$ & 63(8) &39.47 \\
        2$_{2}^{+}$$\rightarrow$2$_{1}^{+}$  & 39(7)  &45.31 & 2$_{2}^{+}$$\rightarrow$0$_{1}^{+}$ & 0.65(11) &1.166 \\
        2$_{3}^{+}$$\rightarrow$2$_{2}^{+}$  &   &  3.42                    & 2$_{3}^{+}$$\rightarrow$0$_{2}^{+}$ &25(7) & 23.23  \\
        0$_{2}^{+}$$\rightarrow$2$_{1}^{+}$  & 0.0121(17)  &4.04 & 0$_{2}^{+}$$\rightarrow$2$_{2}^{+}$  & 99(16) &91.38 \\
        6$_{1}^{+}$$\rightarrow$4$_{1}^{+}$  &  &39.96 & 6$_{1}^{+}$$\rightarrow$4$_{2}^{+}$ &   & 0.477 \\
        2$_{3}^{+}$$\rightarrow$4$_{1}^{+}$  &  & 0.456                     & 2$_{3}^{+}$$\rightarrow$2$_{1}^{+}$ & 2.2(6) &0.65 \\
        4$_{2}^{+}$$\rightarrow$4$_{1}^{+}$  & 24(8)  & 19.26& 4$_{2}^{+}$$\rightarrow$2$_{1}^{+}$ & 0.9(3) &1.85 \\
        4$_{2}^{+}$$\rightarrow$2$_{2}^{+}$  &58(17) &30.85                      & 3$_{1}^{+}$$\rightarrow$4$_{1}^{+}$ & 25(8) &20.59 \\
        3$_{1}^{+}$$\rightarrow$2$_{1}^{+}$  & 1.8(5)  & 0.70& 3$_{1}^{+}$$\rightarrow$2$_{2}^{+}$ & 64(18) &28.21 \\
        0$_{3}^{+}$$\rightarrow$2$_{1}^{+}$  &  &  0.164                    & 0$_{3}^{+}$$\rightarrow$2$_{2}^{+}$ & &  0.093 \\\hline
        $Q_{2^{+}_{1}}$  &-0.38 & -0.045  \\ \hline
    \end{tabular}
    \label{dsafdsfdfd}
\end{table}

\begin{figure}[tbh]
\includegraphics[scale=0.55]{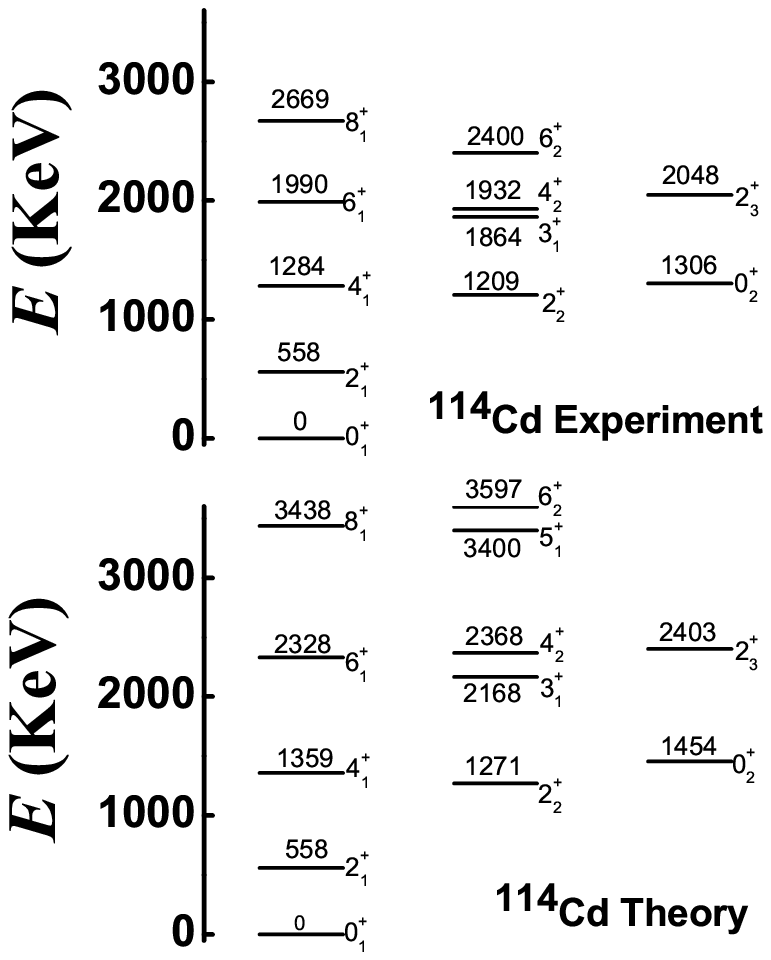}
\caption{Energy spectra of the normal states of $^{114}$Cd (Top: Experiment, Bottom: Theory).}
\end{figure}

\begin{table}[!ht]
    \centering
    \caption{Absolute B(E2) values in W.u. for $E2$ transitions from the low-lying normal states of $^{114}$Cd with effective charge $e=1.41$ (W.u.)$^{1/2}$. The last row is the $Q_{2_{1}^{+}}$ value in eb.}
    \begin{tabular}{llllll}
    \hline
        B(E2)~~~~ & Exp. & Theory~~~~~~&B(E2)~~~~ & Exp. &Theory \\ \hline
        2$_{1}^{+}$$\rightarrow$0$_{1}^{+}$  &31.1(19)     & 31.1        & 4$_{1}^{+}$$\rightarrow$2$_{1}^{+}$ & 62(4) &41.39 \\
        2$_{2}^{+}$$\rightarrow$2$_{1}^{+}$  &22(6) &44.72 & 2$_{2}^{+}$$\rightarrow$0$_{1}^{+}$ & 0.48(6)  &1.72 \\
        2$_{3}^{+}$$\rightarrow$2$_{2}^{+}$  & $<$1.9  & 3.14& 2$_{3}^{+}$$\rightarrow$0$_{2}^{+}$ & 17& 24.51 \\
        0$_{2}^{+}$$\rightarrow$2$_{1}^{+}$  & 0.0026(4)  &  3.65            & 0$_{2}^{+}$$\rightarrow$2$_{2}^{+}$ & 127(16) &89.58 \\
        6$_{1}^{+}$$\rightarrow$4$_{1}^{+}$  & 119(15) &43.54 & 6$_{1}^{+}$$\rightarrow$4$_{2}^{+}$ & & 0.56 \\
        2$_{3}^{+}$$\rightarrow$4$_{1}^{+}$  &  & 0.54                     & 2$_{3}^{+}$$\rightarrow$2$_{1}^{+}$ & & 0.53 \\
        4$_{2}^{+}$$\rightarrow$4$_{1}^{+}$  &  &  20.39                    & 4$_{2}^{+}$$\rightarrow$2$_{1}^{+}$ & &1.60  \\
        4$_{2}^{+}$$\rightarrow$2$_{2}^{+}$  & & 31.92                     & 3$_{1}^{+}$$\rightarrow$4$_{1}^{+}$ & & 20.99 \\
        3$_{1}^{+}$$\rightarrow$2$_{1}^{+}$  &   &0.88                      & 3$_{1}^{+}$$\rightarrow$2$_{2}^{+}$ & &30.72  \\
        0$_{3}^{+}$$\rightarrow$2$_{1}^{+}$  &   & 0.21                     & 0$_{3}^{+}$$\rightarrow$2$_{2}^{+}$ & &0.059  \\\hline
        $Q_{2^{+}_{1}}$  & -0.35  & -0.018 \\  \hline
    \end{tabular}
    \label{dsafdsfdfd}
\end{table}

\begin{figure}[tbh]
\includegraphics[scale=0.55]{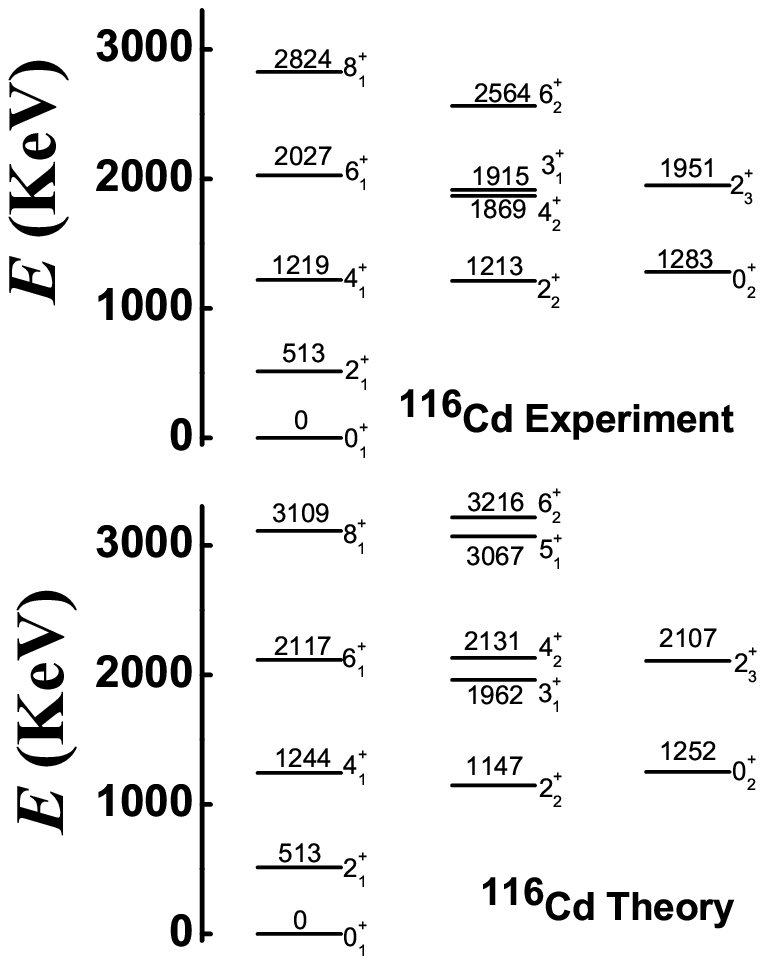}
\caption{Energy spectra of the normal states of $^{116}$Cd (Top: Experiment, Bottom: Theory).}
\end{figure}

\begin{table}[!ht]
    \centering
    \caption{Absolute B(E2) values in W.u. for $E2$ transitions from the low-lying normal states of $^{116}$Cd with effective charge $e=1.61$ (W.u.)$^{1/2}$. The last row is the $Q_{2_{1}^{+}}$ value in eb.}
    \begin{tabular}{llllll}
    \hline
        B(E2)~~~~ & Exp. & Theory~~~~~~&B(E2)~~~~ & Exp. &Theory \\ \hline
        2$_{1}^{+}$$\rightarrow$0$_{1}^{+}$  &33.5(12)   & 33.5                 & 4$_{1}^{+}$$\rightarrow$2$_{1}^{+}$ & 56(14) &43.63 \\
        2$_{2}^{+}$$\rightarrow$2$_{1}^{+}$  & 25(10) & 50.08                  & 2$_{2}^{+}$$\rightarrow$0$_{1}^{+}$ & 1.11(18) & 1.29\\
        2$_{3}^{+}$$\rightarrow$2$_{2}^{+}$  &  &  3.80                    & 2$_{3}^{+}$$\rightarrow$0$_{2}^{+}$ & &25.68  \\
        0$_{2}^{+}$$\rightarrow$2$_{1}^{+}$  &0.79(22)  &4.47                  & 0$_{2}^{+}$$\rightarrow$2$_{2}^{+}$ & 3.0$\times$10$^{4}$(8) &101 \\
        6$_{1}^{+}$$\rightarrow$4$_{1}^{+}$  & 1.1$\times$10$^{2}$$_{-8}^{+4}$ &44.17     & 6$_{1}^{+}$$\rightarrow$4$_{2}^{+}$ & &0.53  \\
        2$_{3}^{+}$$\rightarrow$4$_{1}^{+}$  &    & 0.5                     & 2$_{3}^{+}$$\rightarrow$2$_{1}^{+}$ & & 0.72  \\
        4$_{2}^{+}$$\rightarrow$4$_{1}^{+}$  &  &21.28                      & 4$_{2}^{+}$$\rightarrow$2$_{1}^{+}$ & &2.04  \\
        4$_{2}^{+}$$\rightarrow$2$_{2}^{+}$  & & 34.07                     & 3$_{1}^{+}$$\rightarrow$4$_{1}^{+}$ & &22.76  \\
        3$_{1}^{+}$$\rightarrow$2$_{1}^{+}$  &  & 0.775                     & 3$_{1}^{+}$$\rightarrow$2$_{2}^{+}$ & &31.18  \\
        0$_{3}^{+}$$\rightarrow$2$_{1}^{+}$  &  & 0.181                     & 0$_{3}^{+}$$\rightarrow$2$_{2}^{+}$ & &0.103  \\\hline
        $Q_{2^{+}_{1}}$  &-0.42  & -0.049 \\  \hline
    \end{tabular}
    \label{dsafdsfdfd}
\end{table}

\begin{figure}[tbh]
\includegraphics[scale=0.55]{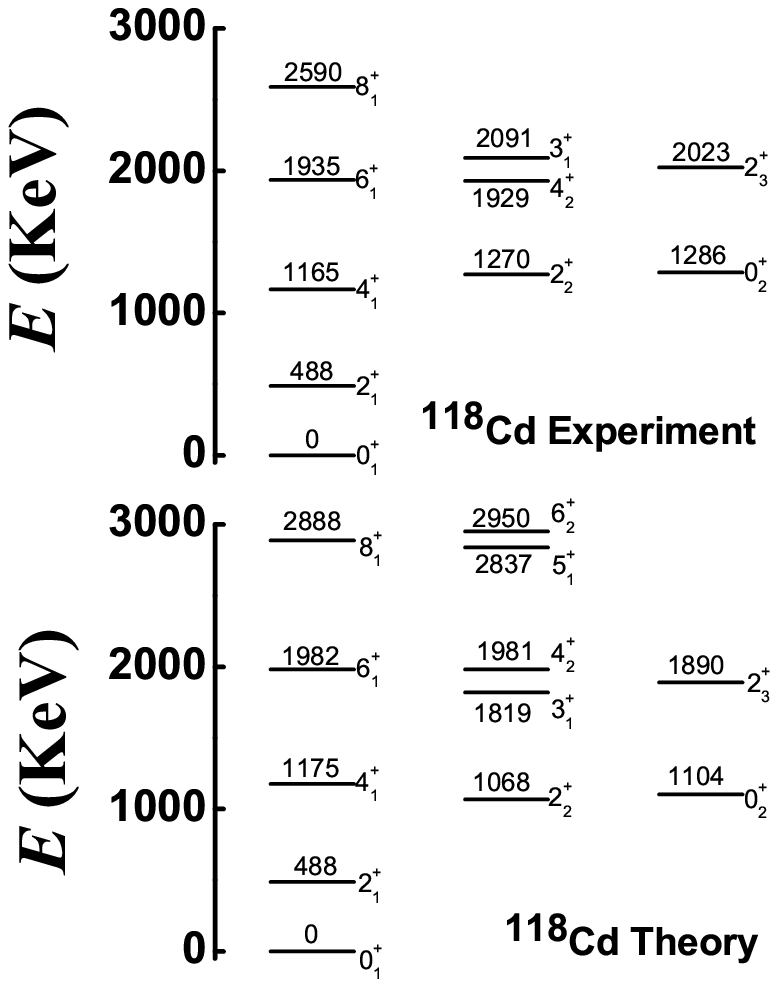}
\caption{Energy spectra of the normal states of $^{118}$Cd (Top: Experiment, Bottom: Theory).}
\end{figure}

\begin{table}[!ht]
    \centering
    \caption{Absolute B(E2) values in W.u. for $E2$ transitions from the low-lying normal states of $^{118}$Cd with effective charge $e=1.76$ (W.u.)$^{1/2}$. The last row is the $Q_{2_{1}^{+}}$ value in eb.}
    \begin{tabular}{llllll}
    \hline
        B(E2)~~~~ & Exp.  & Theory~~~~~~&B(E2)~~~~ & Exp. &Theory\\ \hline
        2$_{1}^{+}$$\rightarrow$0$_{1}^{+}$  &33(3)     & 33                   & 4$_{1}^{+}$$\rightarrow$2$_{1}^{+}$ &$>$61 & 41.61\\
        2$_{2}^{+}$$\rightarrow$2$_{1}^{+}$  &   & 48.8                     & 2$_{2}^{+}$$\rightarrow$0$_{1}^{+}$ & & 1.16  \\
        2$_{3}^{+}$$\rightarrow$2$_{2}^{+}$  &   & 4.35                     & 2$_{3}^{+}$$\rightarrow$0$_{2}^{+}$ & &24.69  \\
        0$_{2}^{+}$$\rightarrow$2$_{1}^{+}$  & 5.3(8)  &5.07 & 0$_{2}^{+}$$\rightarrow$2$_{2}^{+}$ & &102.9   \\
        6$_{1}^{+}$$\rightarrow$4$_{1}^{+}$  &   & 39.84& 6$_{1}^{+}$$\rightarrow$4$_{2}^{+}$ & &0.40  \\
        2$_{3}^{+}$$\rightarrow$4$_{1}^{+}$  &  & 0.37                     & 2$_{3}^{+}$$\rightarrow$2$_{1}^{+}$ & &0.87  \\
        4$_{2}^{+}$$\rightarrow$4$_{1}^{+}$  &  &19.75                      & 4$_{2}^{+}$$\rightarrow$2$_{1}^{+}$ & &2.49  \\
        4$_{2}^{+}$$\rightarrow$2$_{2}^{+}$  & & 32.56                     & 3$_{1}^{+}$$\rightarrow$4$_{1}^{+}$ & &22.86  \\
        3$_{1}^{+}$$\rightarrow$2$_{1}^{+}$  &  & 0.57                     & 3$_{1}^{+}$$\rightarrow$2$_{2}^{+}$ & &28.77  \\
        0$_{3}^{+}$$\rightarrow$2$_{1}^{+}$  &  & 0.087                     & 0$_{3}^{+}$$\rightarrow$2$_{2}^{+}$ & &0.16  \\\hline
        $Q_{2^{+}_{1}}$  &  & -0.089 \\   \hline
    \end{tabular}
    \label{dsafdsfdfd}
\end{table}

\begin{figure}[tbh]
\includegraphics[scale=0.55]{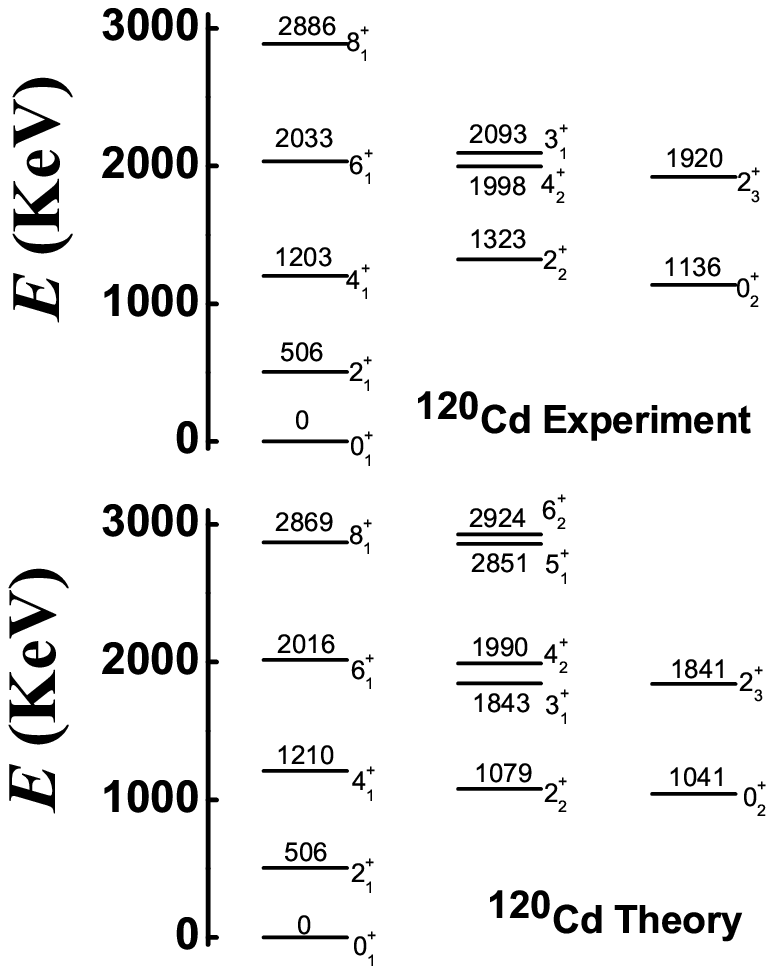}
\caption{Energy spectra of the normal states of $^{120}$Cd (Top: Experiment, Bottom: Theory).}
\end{figure}

\begin{table}[!ht]
    \centering
    \caption{Absolute B(E2) values in W.u. for $E2$ transitions from the low-lying normal states of $^{120}$Cd with effective charge $e=1.0$ (W.u.)$^{1/2}$. The last row is the $Q_{2_{1}^{+}}$ value in eb.}
    \begin{tabular}{llllll}
    \hline
        B(E2)~~~~ & Exp. & Theory~~~~~~&B(E2)~~~~ & Exp. &Theory \\ \hline
        2$_{1}^{+}$$\rightarrow$0$_{1}^{+}$  &   & 8.37                     & 4$_{1}^{+}$$\rightarrow$2$_{1}^{+}$ & & 10.02 \\
        2$_{2}^{+}$$\rightarrow$2$_{1}^{+}$  & &12.23                      & 2$_{2}^{+}$$\rightarrow$0$_{1}^{+}$ & & 0.32  \\
        2$_{3}^{+}$$\rightarrow$2$_{2}^{+}$  &  & 1.27                     & 2$_{3}^{+}$$\rightarrow$0$_{2}^{+}$ &  & 5.94 \\
        0$_{2}^{+}$$\rightarrow$2$_{1}^{+}$  &   & 1.28                     & 0$_{2}^{+}$$\rightarrow$2$_{2}^{+}$ & & 28.43   \\
        6$_{1}^{+}$$\rightarrow$4$_{1}^{+}$  &   &8.75                      & 6$_{1}^{+}$$\rightarrow$4$_{2}^{+}$ &  & 0.01 \\
        2$_{3}^{+}$$\rightarrow$4$_{1}^{+}$  &   & 0.04                     & 2$_{3}^{+}$$\rightarrow$2$_{1}^{+}$ & & 0.325 \\
        4$_{2}^{+}$$\rightarrow$4$_{1}^{+}$  &   & 4.65                     & 4$_{2}^{+}$$\rightarrow$2$_{1}^{+}$ & & 0.75 \\
        4$_{2}^{+}$$\rightarrow$2$_{2}^{+}$  &    &  80                    & 3$_{1}^{+}$$\rightarrow$4$_{1}^{+}$ & &5.92  \\
        3$_{1}^{+}$$\rightarrow$2$_{1}^{+}$  &   & 0.13                     & 3$_{1}^{+}$$\rightarrow$2$_{2}^{+}$ & & 6.48 \\
        0$_{3}^{+}$$\rightarrow$2$_{1}^{+}$  & & 0.0083                     & 0$_{3}^{+}$$\rightarrow$2$_{2}^{+}$ & &0.064  \\\hline
         $Q_{2^{+}_{1}}$  &  & -0.052  \\   \hline
    \end{tabular}
    \label{dsafdsfdfd}
\end{table}

\section{Theoretical fitting of $^{108-120}$Cd}

Previous conclusions are used to fit the $^{108-120}$Cd. \textbf{From Fig. 11, when $\eta=1.0$, $\alpha=1.235$, $\beta=0.0$, $\gamma=-2.0$, $\delta=-4.0$, we can get what we want. There, the spherical-like spectra exist clearly (top graph), and the B(E2) values from the $0_{2}^{+}$ state to the $2_{1}^{+}$ state can be very small (bottom graph). However from Fig. 13, we know that the magnitude of the parameter $\gamma$ can not be large.} \textbf{Thus for the whole Cd isotopes} we focus on the position $\eta=1.0$ and $\alpha=\frac{3N}{2N+3}$, \textbf{which is} the middle point of the transitional region from the U(5) limit to the SU(3) degenerate point \textbf{and the typical position of the spherical-like spectra \cite{wang22}}.\textbf{ The boson number of $^{108-120}$Cd} are $N$=6,7,8,9,8,7,6 respectively. It is interesting that the parameter $\beta=0.0$, $\gamma=-0.5$, $\delta=-8.0$ are also the same for the whole Cd isotopes. \textbf{The magnitude of the parameter $\gamma$ is reduced from -2.0 to -0.5 to make sure the existence of the negative $Q_{2_{1}^{+}}$ value and to reduce the B(E2) values from the $0_{2}^{+}$ state to the $2_{1}^{+}$ state. The parameter $\delta=-8.0$ can show the spherical-like spectra at a good level. The $\hat{C}_{2}^{2}[SU(3)]$ interaction is not needed here.} Let the energy of the $2_{1}^{+}$ state be the same as the experimental value, we choose the fitting parameters $c$ with 688.64 keV, 829.76 keV, 887.39 keV, 902.9 keV, 738.17, 615.38 keV, 550.48 keV for $^{108-120}$Cd. These results show specific regularity. Fig. 14-20 show the low-lying levels of $^{108-120}$Cd and compared with the experimental data \cite{ensdf}. The general features of the energy levels fit quite well. \textbf{It should be stressed that there is no the $0_{3}^{+}$ state at the energy position of previous three phonon level.} The energies of $0_{2}^{+}$ states are somewhat smaller than the experimental ones while the high-spin states are somewhat larger. Apparently, the fitting effect of $^{118,120}$Cd seems better than the lighter ones. The reason for this is that the coupling between the normal states and the intruder states can be ignored for the two nuclei and the normal states are just the spherical-like $\gamma$-soft model.

\textbf{Some deficiencies also need to be further analyzed here. It is hoped that the fitting effect can be further improved when the configuration mixing is considered in next paper. In this article, a single Hamiltonian is used to fit the whole $^{108-120}$Cd. This proves the SU3-IBM, but also makes some of the levels not fit very good enough. We believe that the price is worth it. In further fitting, the parameters will be adjusted more correctly. For $^{108-118}$Cd, the calculated ground-band and $\gamma$-band are systematically stretched as compared to the observed ones. Considering the $^{120}$Cd, this phenomenon is not very serious, so we believe that these bands would be more consistent with the experimental results when the configuration mixing is considered. We expect this to be confirmed by the latter fitting. The fitting also does not account for the inverted energy levels of the $3_{1}^{+}$ and $4_{2}^{+}$ states in $^{118,120}$Cd. This problem is interesting. We notice that in $^{106}$Pd, the two levels are normal \cite{Sona22,Yates}. The inversion phenomenon will be further discussed in future via careful adjustment of the parameters.
}

Absolute B(E2) values for E2 transitions from the low-lying normal states of $^{108-120}$Cd are shown in TABLE II-VIII. The fitting effects are still good. The key quantity is the B(E2) value between the $0_{2}^{+}$ state and the $2_{1}^{+}$ state. In previous paper \cite{wang22}, without other SU(3) higher-order interactions, this value is 12.2 W.u for $^{110}$Cd. In this paper, it is 4.14 W.u.. In $^{112,114}$Cd, these values are nearly zero, while in $^{116}$Cd, it is 0.79 W.u. and larger than the lighter ones. For $^{118}$Cd, the values is 5.3 W.u. (see also Fig. 3). The theoretical fitting result is 5.07 W.u., which fits well (see TABLE VII). This is the target result of this second paper, and its emergence brings about consistency between theory and experiments. Thus it can be suggested that the nearly zero values of $^{112,114}$Cd and the small value of $^{116}$Cd may result from the configuration mixing between the normal states and the intruder states. This will be discussed in future paper.

It should be noticed that the theoretical values $Q_{2_{1}^{+}}$ in $^{108-116}$Cd are much smaller than the experimental data. In previous analysis, we can obtain that the coupling between the normal states and the intruder states in $^{118,120}$Cd can be ignored (this is also supported in \cite{Batchelder12}). So the deviation of $Q_{2_{1}^{+}}$ values in $^{108-116}$Cd may result from configuration mixing. The first intruder states in $^{114}$Cd is the lowest among the whole Cd isotopes, so its theoretical values $Q_{2_{1}^{+}}$  has the largest deviation from the experimental value. If so, the experimental values $Q_{2_{1}^{+}}$ in $^{118,120}$Cd can be predicted at around -0.09 eb, much smaller than the values in $^{108-116}$Cd. We strongly suggest that the \textbf{future} experiment can be able to measure this \textbf{value} and, if it is so, to prove the validity of the SU3-IBM.

\begin{figure}[tbh]
\includegraphics[scale=0.3]{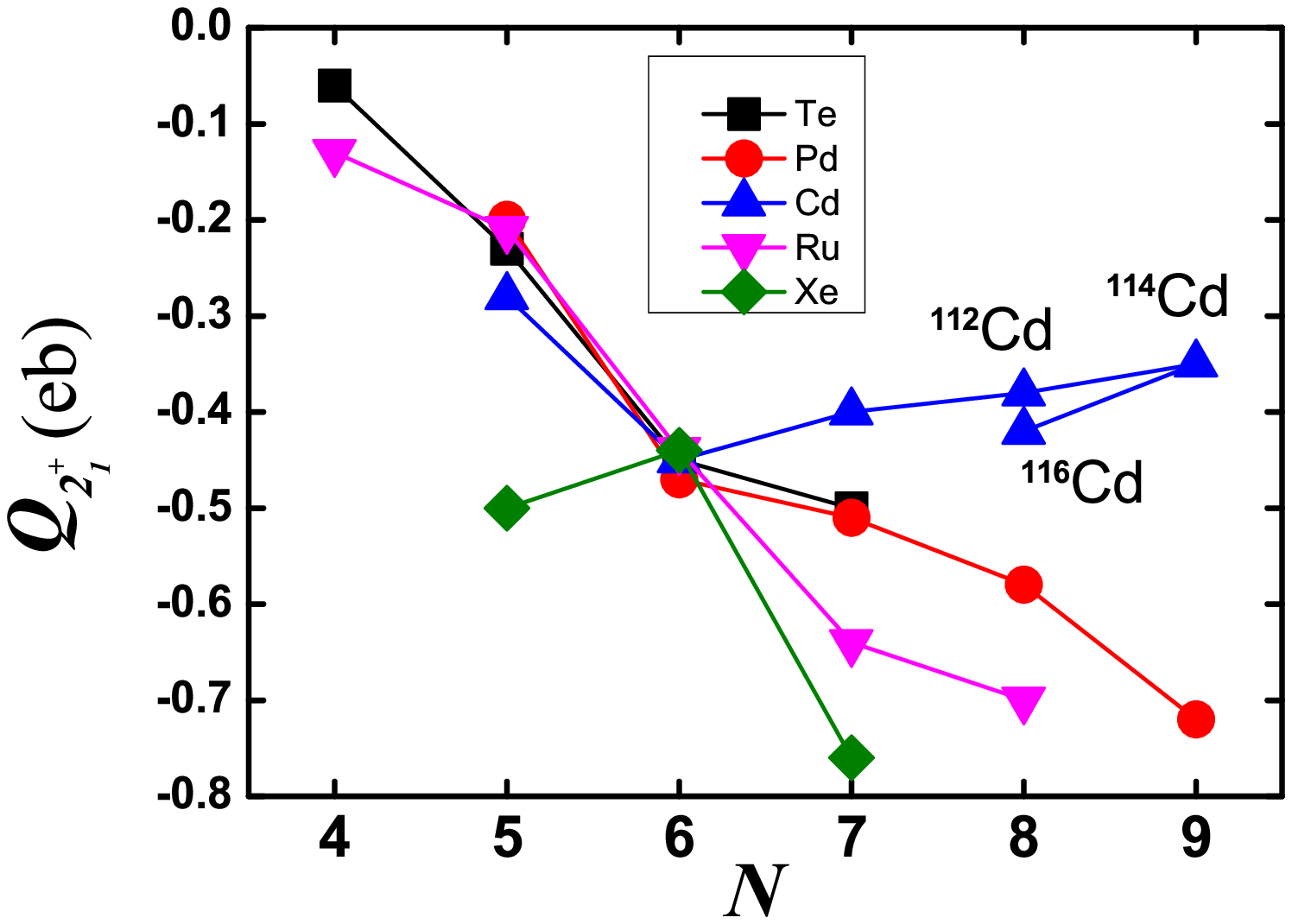}
\includegraphics[scale=0.3]{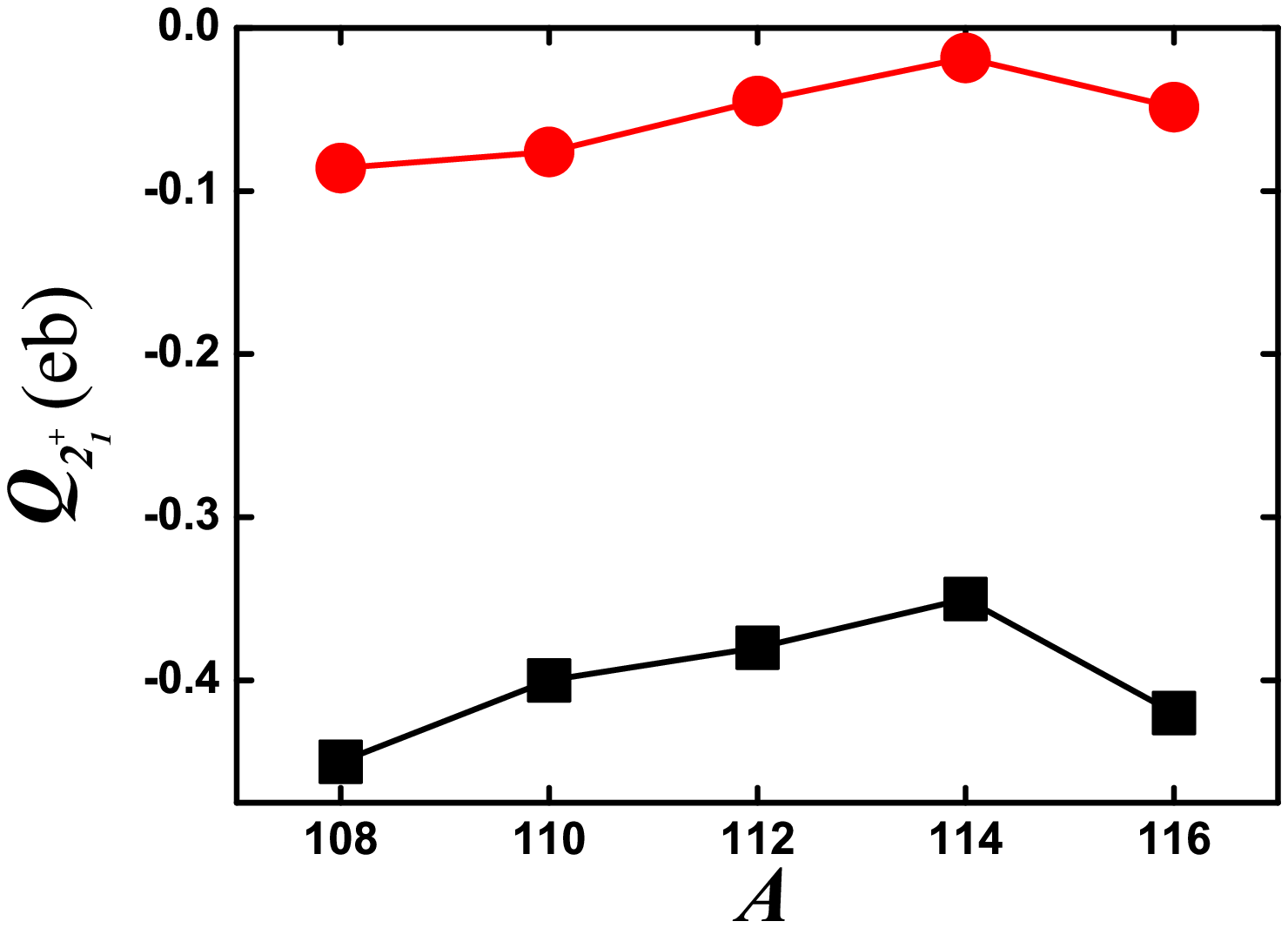}
\caption{Top: the evolutional behaviors of the \textbf{experimental} $Q_{2_{1}^{+}}$ values in Te, Pd, Cd, Ru, Xe nuclei \cite{ensdf}. Bottom: the evolutional behaviors of the theoretical \textbf{(red)} and experimental \textbf{(black)} results for $^{108-116}$Cd.}
\end{figure}

In the top graph of Fig. 21, the evolutional behaviors of the $Q_{2_{1}^{+}}$ values in Te, Pd, Cd, Ru, Xe nuclei are shown. It is clearly seen that the evolutional trend in $^{108-116}$Cd are different from the neighbouring ones, and decreases as $N$ increases. This is an anomalous phenomenon. (To the best of our knowledge, this anomaly is mentioned for the first time in this paper.) Although these theoretical $Q_{2_{1}^{+}}$ values are much smaller than the experimental ones, they reproduce the anomalous trend, see the bottom of Fig. 21. The differences between them comes from the configuration mixing. This result means that both our hypothesis and our results are reasonable.

The theoretical fitting and the practical situation are consistent, and the deficiency comes from the lack of considering the configuration mixing. For $^{108-116}$Cd, the coupling effect, although weak, must be considered, and the resulting deficiency is clear. For $^{118,120}$Cd, the fitting is very reasonable, indicating that the coupling between the normal states and the intruder states can be ignored. We look forward to more experimental studies on $^{118,120}$Cd.

The energy level feature of a nearby nucleus $^{106}$Pd is found to also possess both of these features. $0_{3}^{+}$ state is found to be an intruder state \cite{Sona22}, such that the $0_{4}^{+}$ state is in fact the third $0^{+}$ state of the normal states \cite{Yates}, and its energy is 1.77 times than that of $0_{2}^{+}$ state (see Fig. 6 in \cite{Yates}). The energy level feature at the four phonon level in the spherical-like spectra can be verified. This will be discussed in future. In this paper, we show that the new spherical-like $\gamma$-soft mode can be actually found in the normal states of $^{118,120}$Cd.

\section{Discussions}

\textbf{This is the second paper discussing the Cd puzzle with the SU3-IBM. The first article pointed out that the Cd puzzle implies a new mode of collective motion, which means new possibility for the origin of the collective nature of the nuclei. And in the SU3-IBM, this mode was really found via introducing the SU(3) third-order interaction. In this paper, we continue to examine this idea in preparation for further configuration mixing calculations.}

\textbf{In three aspects, the rationality of the theory is further verified. First, we fit the whole $^{108-120}$Cd using a single Hamiltonian. When the SU(3) higher-order interactions are added, the sphere-like spectra can be described more accurately, which is also the first purpose of this study. Except for some deficiencies, the overall fitting effect of the energy spectra of these isotopes is good at a qualitative level. Especially for the $^{118,120}$Cd, the fitting results agree with the experimental values. For using only the single Hamiltonian, the change of the spectra of the Cd nuclei depends on the boson number $N$. }

\textbf{The second aspect is the B(E2) values from the $0_{2}^{+}$ state to the $2_{1}^{+}$ state. The purpose of this paper is to further reduce this B(E2) value on the results in the first article. The experimental values in $^{112,114}$Cd are nearly zero, which provides a challenge for nuclear structure theories. In this paper, the theory can explain the B(E2) value in $^{118}$Cd at the good level. The theoretical value is 5.07 W.u. while the experimental one is 5.3(8) W.u.. This result illustrates that the nearly zero B(E2) values in $^{112,114}$Cd are due to two reasons. One is that the value of the spherical-like $\gamma$-soft mode itself is relatively small, and the other comes from the configuration mixing. We expect further calculations to confirm this.}

\textbf{The third and most important aspect is to give the anomalous evolution trend of the quadrupole moment of the $2_{1}^{+}$ state $Q_{2_{1}^{+}}$. This result is especially valuable given that only one Hamiltonian is used here. In fact, this is also an unexpected finding. This anomaly is induced by the spherical-like $\gamma$-soft rotation. The differences between the experimental and theoretical values comes from the configuration mixing, which will be verified in the next paper.}

\textbf{Based on recent work on $^{196-204}$Hg \cite{Wang24}, and the work in this paper, we will find that the fit is sensitive to the boson number $N$, which confirms the basic assumption of the IBM. Thus the bosons here are highly like the nucleon pairs. In the actual case, distinguishing the protons and neutrons should be important for understanding the spherical-like spectra. So further generalization to SU3-IBM-2 is needed, and we may find that the Hamiltonian will make differences between the description of protons and neutrons. This will also help us to better understand the evolution of the Cd isotopes and further improve the fitting effect.}

\textbf{The nuclear energy density functionals are also very useful for understanding shape evolution, which is also explored to study the properties of the Cd nuclei \cite{Ro08,Pr12,Nomura18,Go21,Nomura22}. This approach establishes the connection between nuclear force theory and nuclear structure theory. The SU3-IBM has the potential to further improve the understanding of the nuclear forces, so establishing a link between the energy density functional and SU3-IBM is necessary, like the works by Nomura \emph{et al.} \cite{Nomura08,Nomura10,Nomura11,Nomura12}.}

Clearly, these new results provide a key step towards completely solving the spherical nucleus puzzle. From the studies on the SU3-IBM \cite{wang,Wang20,wang22,wang23,Wang24,Zhang22,Zhang22,Zhang22,Zhang24,zhou23,zhou24} and this paper, the SU3-IBM is a more plausible model to describe the low-energy excitation behaviors of realistic nuclei. Ref. \cite{wang22} and this paper stress on the Cd puzzle. These works are even more important because they support denying the existence of spherical nuclei. Although the absence of spherical nuclei has been proposed in Ref. \cite{Garrett08,Garrett10,Garrett12,Wood16,Wood18}, there is no theory to really support this conclusion, especially to answer what this pattern is. In this paper, we show that the SU3-IBM can offer a reasonable answer, and even it fits well in many details.

In the SU3-IBM, two features are not found in previous nuclear structure theories: the higher-order interactions, and the SU(3) symmetry that the interactions have except for the pairing one. In this paper, we show that the SU(3) second-order interaction and the SU(3) third-order interaction both play the same important roles, and other higher-order interactions are also necessary (although small). As we all know, the theories of nuclear structure are based on the mean field perspective, so the residual interactions should be governed by the two-body interactions. But the Cd puzzle breaks down with this idea. In our theory, three-body interaction can play the same role as the two-body one. In Ref. \cite{Zhang12,wang23}, the SU(3) third-order interaction can even play a decisive role. These results appear to conflict with the idea of mean field perspective.

It is clear that the experimental results of the Cd puzzle challenge the mean field perspective, and our results support this challenge. The mean field is present, but for the residual interactions, the two-body interactions do not necessarily play the dominant role. Our current understanding of the mean field is inadequate. Recently  Machleidt and Sammarruca raised a question:``Are the energies typically involved in conventional nuclear structure physics low enough to treat nucleons as structure-less objects?'' \cite{Machleidt24} We conjecture that the Cd puzzle may be related to the medium modifications of the nuclear force \cite{Hen17}. We look for more possible studies on this conjecture.

\section{Conclusions}

Based on previous work, that Cd puzzle means a spherical-like $\gamma$-soft rotation, other SU(3) higher-order interactions are used to understand this new $\gamma$-soft pattern. SU(3) fourth-order interaction $[(\hat{L} \times \hat{Q})^{(1)} \times (\hat{L} \times \hat{Q})^{(1)}]^{(0)}$ can induce a prominent spherical-like spectra except for the $0^{+}$ state at the three phonon level, which is repelled to a higher energy level (at the four phonon level). This quantity also has very little effect on the electric quadrupole moment of the $2_{1}^{+}$ state. SU(3) third-order interaction $\Omega=[\hat{L} \times \hat{Q} \times \hat{L}]^{(0)}$ and the square of SU(3) second-order Casimir operator $\hat{C}_{2}^{2}[SU(3)]$ can reduce the B(E2) value between the $0_{2}^{+}$ state and the $2_{1}^{+}$ state. However they can also reduce the electric quadrupole moment of the $2_{1}^{+}$ state, which poses strict restrictions on the parameters of these two quantities.

It is encouraging that a simple set of parameters can be used to fit the $^{108-120}$Cd, especially $^{118,120}$Cd. The coupling between the normal states and the intruder states in $^{118,120}$Cd can be ignored. This conclusion is also supported by the experimental data, that the intruder states in $^{118,120}$Cd are higher than the ones in $^{108-116}$Cd. For $^{118}$Cd, the B(E2) value between the $0_{2}^{+}$ state and the $2_{1}^{+}$ state fits very well, which is one of the main results of this paper.

This paper predicts that the values of the electric quadrupole moment of the $2_{1}^{+}$ state in $^{118,120}$Cd are about -0.09 eb,  which is expected to need experimental verification.

The normal states in $^{118,120}$Cd are suggested to be typical examples of the spherical-like spectra, and the normal states in $^{106}$Pd and other possible spherical-like spectra will be investigated in future.

These reasonable results imply that the assumption of a spherical-like $\gamma$-soft rotational mode is reasonable and that SU3-IBM can be used to explain the spherical nucleus puzzle. The remaining deficiency on the Cd puzzle should come from configuration mixing, which will be performed in the third paper.

\section{ACKNOWLEDGMENT}

This research is supported by the Educational Department of Jilin Province, China (JJKH20210526KJ). Y. Zhang gratefully acknowledges support from the Natural Science Foundation of China (11875158).

\end{document}